\documentclass[a4paper,11pt]{article}
\pdfoutput=1
\usepackage{jcappub}
\usepackage{afterpage,booktabs,multirow}
\usepackage{graphicx,subfig}
\usepackage{amssymb,amsmath,bm}
\usepackage[usenames,dvipsnames]{xcolor}
\usepackage[capitalise]{cleveref}
\usepackage{expl3,xstring,xparse}
\usepackage{relsize}
\usepackage{color}

\DeclareMathOperator{\BetaFac}{\kappa}
\DeclareMathOperator{\period}{.}
\DeclareMathOperator{\comma}{,}

\newcommand{\changes}[1]{{\color{black}{#1}}}
\newcommand{\finalchanges}[1]{{\color{black}{#1}}}

\title{\boldmath COLA with scale-dependent growth: applications to screened modified gravity models}
\author[a]{Hans A. Winther,}
\author[a]{Kazuya Koyama,}
\author[c,d]{Marc Manera,}
\author[a]{Bill S. Wright,}
\author[b,a]{and Gong-Bo Zhao}
\affiliation[a]{Institute of Cosmology \& Gravitation, Dennis Sciama Building, University of Portsmouth, Portsmouth, PO1 3FX, UK}
\affiliation[b]{National Astronomy Observatories, Chinese Academy of Science, Beijing, 100012, P.R.China}
\affiliation[c]{Centre  for  Theoretical  Cosmology,  Department  of  Applied  Mathematics  and  Theoretical  Physics, Wilberforce Road, Cambridge CB3 0WA, UK}
\affiliation[d]{Kavli Institute for Cosmology, University of Cambridge, Madingley Road, Cambridge CB3 0HA, UK}

\emailAdd{hans.a.winther@gmail.com}

\abstract{We present a general parallelized and easy-to-use code to perform numerical simulations of structure formation using the COLA (COmoving Lagrangian Acceleration) method for cosmological models that exhibit scale-dependent growth at the level of first and second order Lagrangian perturbation theory. For modified gravity theories we also include screening using a fast approximate method that covers all the main examples of screening mechanisms in the literature. We test the code by comparing it to full simulations of two popular modified gravity models, namely $f(R)$ gravity and nDGP, and find good agreement in the modified gravity boost-factors relative to $\Lambda$CDM even when using a fairly small number of COLA time steps.}

\date{\today}

\begin{document}
\maketitle
\flushbottom

\section{Introduction}

One of the main objectives of current and future large scale structure surveys is to test General Relativity (GR) on cosmological scales. GR is so far a perfect fit to observations in the laboratory and in the Solar System \cite{2014LRR....17....4W} and with current and upcoming large scale structure surveys such as Euclid \cite{2015arXiv150104908S}, LSST \cite{2008arXiv0805.2366I}, WFIRST \cite{2013arXiv1305.5422S}, DESI \cite{2016arXiv161100036D}, eBOSS \cite{2016AJ....151...44D} and SKA \cite{godfrey2012} we will be able to place high precision constraints on cosmological scales \cite{2016RPPh...79d6902K, 2015CQGra..32x3001B,doi10}.

One of the possibilities for explaining the late time acceleration of the Universe, if not due to a cosmological constant, is having modifications of gravity \cite{2012PhR...513....1C}. If GR is modified on cosmological scales then some form of screening mechanism is needed to hide the modifications in the Solar System. The existence of a screening mechanism together with the requirement of satisfying the stringent constraints from local tests of gravity, often implies that the most interesting signatures are to be found in the non-linear regime of structure formation. This requires numerical (usually {\it N}-body) simulations in order to compute theoretical predictions which can then be compared with data and used to place constraints on modified gravity models.

{\it N}-body simulations of modified gravity have been around for about a decade now and in that time several different codes have been created \cite{2008PhRvD..78l3523O,2009PhRvD..80j4005C,2009PhRvD..80d3001S,2009PhRvD..80f4023K,2011PhRvD..83b4007L,2011PhRvD..83d4007Z,2012JCAP...01..051L,2013MNRAS.436..348P,2014A&A...562A..78L} to produce high-resolution simulations for a wide variety of models. A recent code-comparison project of such codes \cite{2015MNRAS.454.4208W} demonstrated agreement to the $1\%$ level deep into the non-linear regime (e.g. $k\sim 5 h/$Mpc for the power-spectrum).

Originally, modified gravity simulations were much slower than simulations of $\Lambda$CDM (typically by a factor of $5-20$ depending on model) due to having to solve complicated, highly non-linear partial differential equations. However, recently some very interesting approaches have been proposed to speed up such simulations making them only a factor of $\sim 1.5-2$ times slower than a corresponding $\Lambda$CDM simulation without sacrificing much accuracy \cite{2016arXiv161109375B,2015JCAP...12..059B,2015PhRvD..91l3507W}.

However, for many purposes there is still a need for even faster methods, even for the case of $\Lambda$CDM. For example, to study weak lensing and galaxy clustering in current and future large structure surveys there is a need to generate huge ensembles of mock halo catalogs needed \changes{to model the observables and their covariances.}

Several fast, approximate methods for such purposes have been proposed over the last decade like PINOCCHIO (PINpointing Orbit-Crossing Collapsed HIerarchical Objects) \cite{2002MNRAS.331..587M}, which recently has been extended to including massive neutrinos and modified gravity \cite{2017JCAP...01..008R},  as well as Peak-Patch \cite{1996ApJS..103....1B}, PTHalos \cite{2002MNRAS.329..629S,2013MNRAS.428.1036M,2015MNRAS.447..437M}, QPM (Quick Particle Mesh) \cite{2014MNRAS.437.2594W}, PATCHY (PerturbAtion Theory Catalog generator of
Halo and galaxY distributions) \cite{2014MNRAS.439L..21K}, HALOGEN \cite{2015MNRAS.450.1856A} and COLA (COmoving Lagrangian Acceleration) \cite{2013JCAP...06..036T,2016arXiv161206469V,2015arXiv150207751T,2016MNRAS.459.2118K,2015JCAP...06..015L,2016MNRAS.459.2327I,2017arXiv170400920M}. The COLA method, which is the one we will work on extending in this paper, works by placing the {\it N}-body particles in a frame that is co-moving with observers following the path dictated by Lagrangian perturbation theory. This means that we can take fairly large time steps in the {\it N}-body code without loosing accuracy on large scale at the expense of sacrificing accuracy on small scales. This property makes this method much faster, \changes{typically by a factor $O(100-1000)$}, than conventional {\it N}-body simulations. In the limit where the number of steps we use gets larger and larger the method will converge to the result of a standard {\it N}-body simulation (with the same simulation parameters). \changes{A comprehensive study of the accuracy of COLA with respect to the simulations parameters can be found in \cite{2015MNRAS.452..686C,2016MNRAS.459.2327I,2017arXiv170400920M}.}

Recently, in \cite{2016arXiv161206469V} an adaptation of the COLA approach for chameleon and symmetron modified gravity models was proposed and shown to work very well. It was shown that even though COLA overestimates the halo mass function for $\Lambda$CDM, the relative changes with respect to $\Lambda$CDM remains accurate.

In this paper we present a code, {\tt{MG-PICOLA}}\footnote{The code can be found at https://github.com/HAWinther/MG-PICOLA-PUBLIC The original {\tt{L-PICOLA}} code can be found at https://github.com/CullanHowlett/l-picola}, based on the publicly available {\tt{L-PICOLA}} code \cite{2015A&C....12..109H}, that allows us to perform numerical simulation of structure formation for general theories that exhibit scale-dependent growth using the COLA approach. The code computes the second order Lagrangian displacement-fields for these theories and also includes general methods to take into account the all important screening effect in modified gravity theories. We have implemented three types of common screening mechanisms: potential (chameleon \cite{2004PhRvD..69d4026K,2007PhRvD..75f3501M}, symmetron \cite{2010PhRvL.104w1301H,2008PhRvD..77d3524O,2005PhRvD..72d3535P} etc.), gradient (k-Mouflage \cite{2009IJMPD..18.2147B}) and density (the Vainhstein mechanism \cite{1972PhLB...39..393V}; DGP, Galileon models). We have implemented often studied models like $f(R)$ and DGP together with a general $\{m(a),\beta(a)\}$ parameterization \cite{2012PhRvD..86d4015B,2012PhLB..715...38B} of modified gravity models with chameleon-like screening. Our approach is therefore able to cover most of the popular models that have been proposed in the literature.

The structure of this paper is as follows: in Sec.~\ref{sec:colalcdm} and Sec.~\ref{sec:lptlcdm} we give an overview of the COLA method and Lagrangian perturbation theory in $\Lambda$CDM before we derive the general equations for models with scale-dependent growth in Sec.~\ref{sec:lptmg}. In Sec.~\ref{sec:mgmodels} we give a brief description of the modified gravity models we consider in this paper before describing the approximate method we use to include (three types of) screening mechanisms in Sec.~\ref{sec:screening}. In Sec.~\ref{sec:results} we show results of our method before concluding in Sec.~\ref{sec:conclusion}. Details about the code implementation can be found in the Appendix.

\changes{Unless stated otherwise all time-derivatives are with respect to the super-comoving time-coordinate $\tau$ defined by ${\rm d}\tau = \frac{{\rm d}t}{a^2}$ and $\BetaFac \equiv 4\pi G\overline{\rho} a^4 = \frac{3}{2}\Omega_mH_0^2a$.}

\section{The COLA approach to non-linear structure formation}\label{sec:colalcdm}
In a typical cold-dark-matter {\it N}-body simulation we solve the equations\footnote{Note that the gravitational potential here is $a^2$ times the conventional gravitational potential (the metric perturbation).}
\begin{align}\label{eq:nbody}
\frac{d^2\vec{x}}{d\tau^2} = -\vec{\nabla_{\bf x}}\Phi_N\comma\\
\nabla_{\bf x}^2\Phi_N = 4\pi G\overline{\rho}a^4\delta \equiv \BetaFac\delta\comma
\end{align}
where ${\rm d}\tau = \frac{{\rm d}t}{a^2}$ and $\delta = \frac{\rho}{\overline{\rho}}-1$ is the matter density contrast which is computed from the particle positions $\vec{x}$. To get high accuracy we need to take very small time steps which makes such computations expensive to run. However if one is only interested in scales where the evolution of the density field is quasi-linear, but where non-linear effects are still important to get accurate results, then there is a useful trick to speed up such simulations. Instead of solving for the positions of the particles $\vec{x}$, in the COLA approach \cite{2013JCAP...06..036T} we solve for the perturbation about the path $\vec{x}_{\rm LPT}$ predicted from second order Lagrangian perturbation theory. Taking $\vec{x} = \vec{\delta x} + \vec{x}_{\rm LPT}$ gives us the geodesic equation
\begin{align}
\frac{d^2\vec{\delta x}}{d\tau^2} = -\vec{\nabla_{\bf x}}\Phi_N - \frac{d^2\vec{x}_{\rm LPT}}{d\tau^2}\comma
\end{align}
which is solved as the coupled system
\begin{align}
\frac{d\vec{\delta v}}{d\tau} &= -\vec{\nabla_{\bf x}}\Phi_N - \frac{d^2\vec{x}_{\rm LPT}}{d\tau^2}\comma\\
\frac{d\vec{\delta x}}{d\tau} &= \vec{\delta v}\comma
\end{align}
typically using a Leapfrog integrator. Since the large scale evolution of the particles will be close to that of Lagrangian perturbation theory it means we are able to take much larger time steps in the simulations making it much faster than a standard {\it N}-body simulations. This of course comes at the expense of accuracy on small scales.

\section{Lagrangian perturbation theory for $\Lambda$CDM}\label{sec:lptlcdm}

In Lagrangian perturbation theory (LPT; see \cite{1996dmu..conf..565B} for a review) the \changes{comoving position} of a particle $\vec{x}$ is written in terms of its initial position $\vec{q}$ and a displacment field $\vec{\Psi}$ as $\vec{x} = \vec{q} + \vec{\Psi}(\vec{q},\tau)$. The geodesic equation (\ref{eq:nbody}) can be written
\begin{align}\label{eq:geodesicdispfield}
\frac{d^2}{d\tau^2}\vec{\Psi}_{i,i} - \vec{\Psi}_{j,i}\frac{d^2}{d\tau^2}\vec{\Psi}_{i,j} = -\nabla_{\bf x}^2\Phi_N = -\BetaFac \delta\comma
\end{align}
where $\BetaFac = 4\pi G\overline{\rho} a^4 = \frac{3}{2}\Omega_mH_0^2a$ \changes{and $\vec{\Psi}_{i,j}\equiv \frac{d\vec{\Psi}_i}{d\vec{q}_j}$}. We expand the displacement field in a perturbation series $\vec{\Psi} = \epsilon\vec{\Psi}^{(1)} + \epsilon^2\vec{\Psi}^{(2)} + \ldots$ and since $\vec{\Psi}$ is assumed to be curl-free we can write $\vec{\Psi}^{(i)} = \vec{\nabla_{\bf q}}\phi^{(i)}$ where $\phi^{(i)}$ is a scalar field. The density contrast $\delta = \left|\frac{\partial (x,y,z)}{\partial (q_x,q_y,q_z)}\right|^{-1} - 1 = \epsilon\delta^{(1)} + \epsilon^2\delta^{(2)} + \ldots$ can be written in terms of the displacement-field order by order as
\begin{align}
\delta^{(1)} &= - \vec{\Psi}^{(1)}_{i,i}\comma\\
\delta^{(2)} &= -\vec{\Psi}^{(2)}_{i,i} + \frac{1}{2}((\vec{\Psi}^{(1)}_{i,i})^2+(\vec{\Psi}^{(1)}_{i,j})^2)\period
\end{align}
\subsection{First order: 1LPT}
To first order the equations above gives us
\begin{align}
\left(\frac{d^2}{d\tau^2} - \BetaFac\right)\nabla_{\bf q}^2\phi^{(1)} = 0\comma
\end{align}
and $\nabla_{\bf q}^2\phi^{(1)}(\vec{q},\tau_{\rm ini}) = -\delta^{(1)}(\vec{q},\tau_{\rm ini})$ is given by the initial conditions. This can be factorized as $\phi^{(1)}(\vec{q},\tau) = D_1(\tau)\phi^{(1)}(\vec{q},\tau_{\rm ini})$ where the growth-factor $D_1$ only depends on time and satisfies the simple ODE
\begin{align}
\frac{d^2D_1}{d\tau^2} - \BetaFac D_1 = 0\period
\end{align}
The initial conditions are set such that $D_1^{\rm ini} = 1$ and $\frac{dD_1^{\rm ini}}{d\tau} = \left(\frac{1}{a}\frac{da}{d\tau}\right)_{\tau=\tau_{\rm ini}}$ corresponding to the growing mode in a matter dominated universe (Einstein-de Sitter).

The displacement-field at any time satisfies $\vec{\Psi}^{(1)}(\vec{q},\tau) = D_1(\tau)\vec{\Psi}^{(1)}(\vec{q},\tau_{\rm ini})$ which means that in a numerical simulation we need only compute $\vec{\Psi}^{(1)}$ once at the initial time and store this for each particle and then use the growth-factor to compute it at any subsequent time.

\subsection{Second order: 2LPT}
To second order Eq.~(\ref{eq:geodesicdispfield}) gives us
\begin{align}
\left(\frac{d^2}{d\tau^2} - \BetaFac\right)\nabla_{\bf q}^2\phi^{(2)} = - \frac{\BetaFac}{2}\left[ (\nabla_{\bf q}^2\phi^{(1)})^2 - (\nabla_{\textbf{q}_i}\nabla_{\textbf{q}_j}\phi^{(1)})^2\right]\period
\end{align}
Again we can separate this as $\phi^{(2)}(\vec{q},\tau) = D_2(\tau)\phi^{(2)}(\vec{q},\tau_{\rm ini})$ where
\begin{align}
\frac{d^2D_2}{d\tau^2} - \BetaFac D_2 = -\BetaFac D_1^2\period
\end{align}
For an Einstein-de Sitter Universe the physically relevant solution has $D_2 = -\frac{3}{7}D_1^2$ so the initial conditions are taken to be $D_2^{\rm ini} = -\frac{3}{7}$ and $\frac{dD_2^{\rm ini}}{d\tau} = -\frac{6}{7}\left(\frac{1}{a}\frac{da}{d\tau}\right)_{\tau=\tau_{\rm ini}}$.

The initial field $\phi^{(2)}(\vec{q},\tau_{\rm ini})$ satisfies
\begin{align}\label{eq:lcdm2lpt}
\nabla_{\bf q}^2\phi^{(2)} = \frac{1}{2}\left[(\nabla_{\bf q}^2\phi^{(1)})^2 - (\nabla_{\textbf{q}_i}\nabla_{\textbf{q}_j}\phi^{(1)})^2\right]\comma
\end{align}
which is easy to compute numerically using Fourier transforms and again we only need to compute it once and store the corresponding displacement-vector with each particle.

\section{Lagrangian perturbation theory with scale-dependent growth}\label{sec:lptmg}
\changes{
For theories where the growth-factor is scale-dependent the situation becomes a little more complicated than in $\Lambda$CDM. We will here consider a general, second order parametrization of the gravitational potential in Fourier space \cite{2016JCAP...08..032B}
\begin{align}\label{eq:eulerian}
\mathcal{F}_{\textbf{x}}[\nabla_{\textbf{x}}^2\Phi](\vec{k},a) &= \BetaFac \mu(k,a)\delta^{\rm E}(\vec{k},a) \nonumber\\+ &a^4H^2\int\frac{{\rm d}^3k_1{\rm d}^3k_2}{(2\pi)^3}\delta^{(1)}(\vec{k_1},a)\delta^{(1)}(\vec{k_2},a)\gamma_2^{\rm E}(\vec{k},\vec{k_1},\vec{k_2},a)\period
\end{align}
where $\delta^E(\vec{k},a) \equiv \mathcal{F}_{\textbf{x}}[\delta(\vec{x},a)](\vec{k})$. In Eq.~(\ref{eq:geodesicdispfield}) we need the Fourier transform of $\nabla^2_{\textbf{x}}\Phi$ with respect to the Lagrangian coordinate $\vec{q}$ and in terms of the density contrast $\delta(\vec{k},a) = \mathcal{F}_{\textbf{q}}[\delta(\vec{x},a)]$. To first order there is no difference as $\delta(\vec{k},a) = \delta^E(\vec{k},a)$, but performing this transformation to second order gives us
\begin{align}
\mathcal{F}_{\textbf{q}}[\nabla_{\textbf{x}}^2\Phi](\vec{k},a) = \BetaFac\mu(k,a)\delta(\vec{k},a) + a^4H^2 \int \frac{{\rm d}^3k_1{\rm d}^3k_2}{(2\pi)^3}\delta^{(1)}(\vec{k_1},a)\delta^{(1)}(\vec{k_2},a) \times\nonumber\\
\times\left[\gamma_2^E(\vec{k},\vec{k_1},\vec{k_2},a) + \frac{3}{2}\Omega_m(a)\left[\mu(k,a) - \mu(k_1,a)\right]\frac{\vec{k_1}\cdot \vec{k_2}}{k_2^2}\right]\comma\nonumber\\
= \BetaFac\mu(k,a)\delta(\vec{k},a) + a^4H^2 \int \frac{{\rm d}^3k_1{\rm d}^3k_2}{(2\pi)^3}\delta^{(1)}(\vec{k_1},a)\delta^{(1)}(\vec{k_2},a) \gamma_2(\vec{k},\vec{k_1},\vec{k_2},a)\comma
\end{align}
where we have defined $\gamma_2 = \gamma_2^{\rm E} +  \frac{3}{2}\Omega_m(a)\left[\mu(k,a) - \mu(k_1,a)\right]\frac{\vec{k_1}\cdot \vec{k_2}}{k_2^2}$. The second term in $\gamma_2$\footnote{This term was first pointed out in \cite{2017arXiv170510719A}.} is seen to vanish if $\mu(k,a) = \mu(a)$ (like in nDGP and also in $\Lambda$CDM where $\mu\equiv 1$) or when $\vec{k_1}\cdot \vec{k_2} = 0$. We will later see examples of $\mu$ and $\gamma_2$ for some selected modified gravity theories. 

In the following sections the Fourier transforms we use are all with respect to the Lagrangian position $\vec{q}$.
}

\subsection{First order: 1LPT}

Unlike in $\Lambda$CDM we can no longer separate time and space, however we can separate time for each Fourier mode. Going to Fourier space Eq.~(\ref{eq:geodesicdispfield}) gives us
\begin{align}
\left(\frac{d^2}{d\tau^2} - \BetaFac\mu(k,a) \right)\phi^{(1)}(\vec{k},\tau) = 0\comma
\end{align}
which allows us to make the split $\phi^{(1)}(\vec{k},\tau) = D_1(k,\tau)\phi^{(1)}(\vec{k},\tau_{\rm ini})$ where the growth-factor satisfies
\begin{align}
\frac{d^2D_1}{d\tau^2} - \BetaFac\mu(k,a) D_1 = 0\comma
\end{align}
with initial conditions $D_1(\tau_{\rm ini}) = 1$ and $\frac{dD_1^{\rm ini}}{d\tau} = \left(\frac{1}{a}\frac{da}{d\tau}\right)_{\tau=\tau_{\rm ini}}$.

In our {\tt{L-PICOLA}} implementation we compute and store the initial displacement-field in Fourier space and then at every time-step when we need the displacement-vector we compute it by multiplying by the growth-factor (or the time-derivatives of the growth-factor depending on what we need) and performing Fourier transforms.

\subsection{Second order: 2LPT}

To second order we expand $\phi^{(2)}$ (in a way that will become clear later) as
\begin{align}\label{eq:psi2lptintegral}
\phi^{(2)}(\vec{k},\tau) = -\frac{1}{2k^2}\int\frac{{\rm d}^3k_1{\rm d}^3k_2}{(2\pi)^3}\delta_D(\vec{k}-\vec{k}_{12}) \nonumber\\ \delta^{(1)}(\vec{k_1},\tau_{\rm ini})\delta^{(1)}(\vec{k_2},\tau_{\rm ini})D_2(\vec{k},\vec{k_1},\vec{k_2},\tau)\comma
\end{align}
where $\delta^{(1)}$ corresponds to the initial density field. In this form Eq.~(\ref{eq:geodesicdispfield}) becomes
\begin{align}\label{eq:fofrexact}
\frac{d^2D_2}{d\tau^2} &- \BetaFac\mu(k,a) D_2 = -\BetaFac\mu(k,a) D_1(k_1,\tau)D_1(k_2,\tau)\nonumber\\
&\times \left(1 - \changes{\left(\frac{2\mu(k_1,a) - \mu(k,a)}{\mu(k,a)}\right)}\frac{(\vec{k_1}\cdot\vec{k_2})^2}{k_1^2k_2^2} + \frac{2a^4H^2}{\BetaFac\mu(k,a)}\gamma_2(\vec{k},\vec{k}_1,\vec{k}_2,a)\right)\comma
\end{align}
with initial conditions
\begin{align}
D_2^{\rm ini} &= -\frac{3}{7}\left(1 - \frac{(\vec{k_1}\cdot\vec{k_2})^2}{k_1^2k_2^2}\right)\comma\\
\frac{dD_2^{\rm ini}}{d\tau} &= -\frac{6}{7}\left(1 - \frac{(\vec{k_1}\cdot\vec{k_2})^2}{k_1^2k_2^2}\right)\left(\frac{1}{a}\frac{da}{d\tau}\right)_{\tau=\tau_{\rm ini}}\period
\end{align}
In most cases $\gamma_2$ is only a function of the wavenumber norms $k,k_1,k_2$ in addition to the dot-product $\vec{k_1}\cdot \vec{k_2}$, and since the $\delta_D$ function in the integral for $D_2$ enforces $\vec{k_2} = \vec{k} - \vec{k_1}$ it becomes a three-dimensional problem, i.e. we only need to solve it for all relevant combinations of $k$, $k_1$ and $\cos\theta \equiv \frac{\vec{k_1}\cdot \vec{k_2}}{k_1k_2}$ that correspond to a valid triangle in Fourier space.

Evaluating the integral in Eq.~(\ref{eq:psi2lptintegral}) at each time-step, without being able to rely on fast Fourier transforms, is going to ruin the speed of the COLA approach. We therefore settle on an approximation for this term. We define $\phi^{(2)}(\vec{k},\tau) = \hat{D}_2(k,\tau)\phi^{(2)}(\vec{k},\tau_{\rm ini})$ where
\begin{align}\label{eq:phi2approx}
\phi^{(2)}(\vec{k},\tau_{\rm ini}) = -\frac{1}{2k^2}\int\frac{{\rm d}^3k_1{\rm d}^3k_2}{(2\pi)^3}\delta_D(\vec{k}-\vec{k}_{12}) \nonumber\\
\delta^{(1)}(\vec{k_1},\tau_{\rm ini})\delta^{(1)}(\vec{k_2},\tau_{\rm ini})\left(1 - \frac{(\vec{k_1}\cdot\vec{k_2})^2}{k_1^2k_2^2}\right)\comma
\end{align}
which is nothing but the Fourier space version of Eq.~(\ref{eq:lcdm2lpt}) for $\Lambda$CDM and
\begin{align}\label{eq:fofrapprox}
\frac{d^2\hat{D}_2}{d\tau^2} - \BetaFac \mu(k,a) \hat{D}_2 = -\BetaFac \mu(k, a) D_1^2(k,a)\times\nonumber\\
\times\left(1 + \frac{2a^4H^2}{\BetaFac \mu}\gamma_2(k,k/\sqrt{2},k/\sqrt{2},a)\right)\comma
\end{align}
with initial conditions $\hat{D}_2^{\rm ini} = -\frac{3}{7}$ and
$\frac{d\hat{D}_2^{\rm ini}}{d\tau} = -\frac{6}{7}\left(\frac{1}{a}\frac{da}{d\tau}\right)_{\tau=\tau_{\rm ini}}$.

\begin{figure*}
\begin{center}
\includegraphics[width=0.9\textwidth]{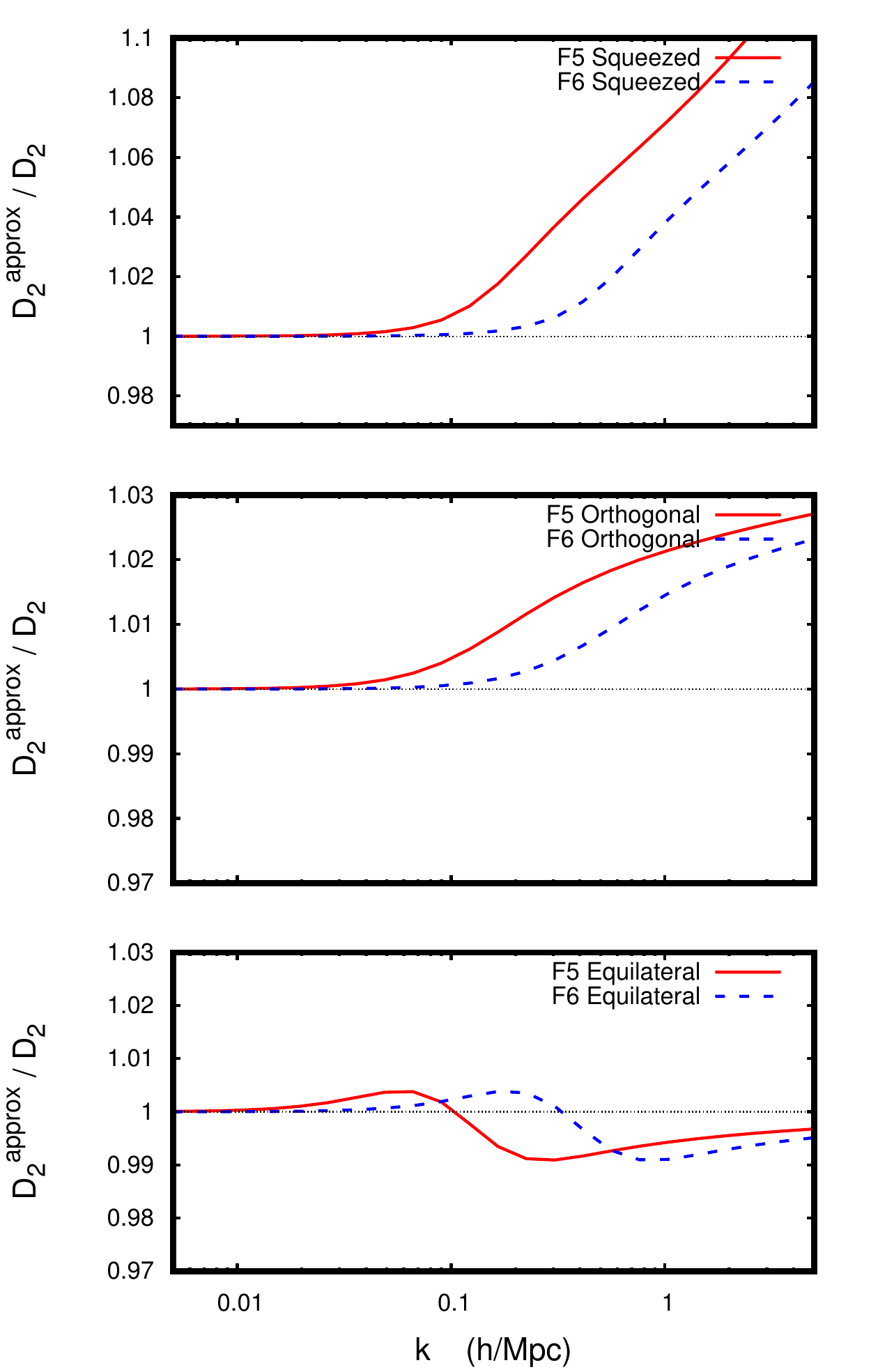}
\caption{The ratio of $D_2(k,k_1,k_2,\cos\theta,a=1)$ to the approximation $D_2(k,a=1)(1-\cos^2\theta)$ for three different triangle configurations; equilateral $k=k_1=k_2$, orthogonal $k_1=k_2=k/\sqrt{2}$ and squeezed $k = k_1$ with $k_2 \approx 0$. Here F5 (F6) refers to a Hu-Sawicky $f(R)$ model with $n=1$ and $|f_{R0}| = 10^{-5}$ ($|f_{R0}| = 10^{-6}$).}
\label{fig:fofrapprox}
\end{center}
\end{figure*}

If $\gamma_2 = 0$ ($\Lambda$CDM) then the equation above is exact. Another case we can do exactly is when $\gamma_2 = f(a)\left(1 - \cos^2\theta\right)$ and\footnote{We must require $\mu$ to be independent of scale in order to put $D_1(k_1,a)D_1(k_2,a) \equiv D_1^2(a)$} $\mu(k,a) = \mu(a)$ like in nDGP. Here the angular dependence of the $\gamma_2$ term is the same as the other term in Eq.~(\ref{eq:fofrexact}) and we can factor out $\left(1 - \cos^2\theta\right)$ to get $D_2(k_1,k_2,k,a) = (1-\cos^2\theta)\hat{D}_2(a)$ where
\begin{align}
\frac{d^2\hat{D}_2}{d\tau^2} - \BetaFac \mu(a) \hat{D}_2 = -\BetaFac \mu(a) D_1^2(k,a)\left(1 + \frac{2a^4H^2}{\BetaFac \mu(a)} f(a)\right)\period
\end{align}
The choice of arguments for $\gamma_2$ in our approximation above is chosen such that it gives the correct equation for the triangle configurations of $\vec{k},\vec{k_1},\vec{k_2}$ giving rise to most of the weight in the integral Eq.~(\ref{eq:psi2lptintegral}). To get an idea about how good this approximation is, in Fig.~\ref{fig:fofrapprox} we show the ratio of our approximation\footnote{We multiply our approximation by $(1-\cos^2\theta)$ when comparing this to $D_2(k,k_1,k_2,a=1)$ as this is the equivalent expression for $\Lambda$CDM. This can be seen from comparing Eq.~(\ref{eq:psi2lptintegral}) to Eq.~(\ref{eq:phi2approx}).} Eq.~(\ref{eq:fofrapprox}) to $D_2(k,k_1,k_2,a=1)$ in Eq.~(\ref{eq:fofrexact}) for different Fourier space triangle configurations of $\vec{k} = \vec{k}_1 + \vec{k}_2$. For the orthogonal and equilateral cases these agree to $\sim 1-2 \%$ up to $k=5h/$Mpc for the models F5 and F6 (defined below) while for the squeezed triangle configuration the difference can be up to $10\%$ for $k\gtrsim 1h/$Mpc.

\section{Modified gravity models}\label{sec:mgmodels}

In this section we give a brief overview of the two modified gravity models we are using in this paper focusing on the equations that are needed for our COLA implementation. For a more thorough review of these models, and modified gravity in general, see \cite{2012PhR...513....1C}.

\subsection{$f(R)$ gravity}
For $f(R)$ gravity \cite{2010LRR....13....3D} the growth of linear perturbations is determined by
\begin{align}
\mu(k,a) = 1 + \frac{1}{3}\frac{k^2}{k^2 + a^2m^2(a)}\comma
\end{align}
where $m(a)$ depends on the model in question. For the ($n=1$) Hu-Sawicky model \cite{2007PhRvD..76f4004H}, which is the $f(R)$ model we will consider in this paper, $m(a)$ is given by
\begin{align}\label{eq:mofahus}
m^2(a) = \frac{1}{3f_{RR}(a)} = \frac{H_0^2(\Omega_m + 4\Omega_\Lambda)}{2|f_{R0}|} \left(\frac{\Omega_m a^{-3} + 4\Omega_\Lambda}{\Omega_m + 4\Omega_\Lambda}\right)^{3}\period
\end{align}
\changes{where $f_R(a) \equiv \left.\frac{df(R)}{dR}\right|_{R = R(a)}$, $f_{R0} = f_R(a=1)$ and $f_{RR}(a) \equiv \left.\frac{d^2f(R)}{dR^2}\right|_{R = R(a)}$}. The field in the cosmological background satisfies
\begin{align}
f_R(a) = f_{R0}\left(\frac{\Omega_m + 4\Omega_\Lambda}{\Omega_ma^{-3} + 4\Omega_\Lambda}\right)^{2}\period
\end{align}
The $\gamma_2^E$ term is given by \cite{2016JCAP...08..032B}
\begin{align}
\gamma_2^{\rm E} &= -\frac{9\Omega_{m}^2}{48a^6|f_{R0}|^2}\left(\frac{k}{aH}\right)^2
\times \frac{(\Omega_m a^{-3} + 4\Omega_\Lambda)^5}{(\Omega_{m} +4\Omega_\Lambda)^4}\frac{1}{\Pi(k,a)\Pi(k_1,a)\Pi(k_2,a)}\comma\\
\end{align}
where
\begin{align}
\Pi(k,a) = \left(\frac{k}{aH_0}\right)^2 + \frac{(\Omega_ma^{-3} + 4\Omega_\Lambda)^3}{2|f_{R0}|(\Omega_m + 4\Omega_\Lambda)^2}\period
\end{align}

\subsection{nDGP gravity}

In nDGP we have a $\Lambda$CDM background expansion, but with  modified growth of perturbations. The growth of linear perturbations are determined by
\begin{align}
\mu(k, a) &= 1 + \frac{1}{3\beta_{\rm DGP}(a)}\comma\\
\beta_{\rm DGP}(a) &= 1 + 2r_cH(a)\left(1 + \frac{\dot{H}}{3H^2}\right)\comma
\end{align}
and $\gamma_2^E$ is given by \cite{2016JCAP...08..032B}
\begin{align}
\gamma_2^{\rm E} = -\left(\frac{H_0}{H}\right)^2\frac{(r_cH_0)^2\Omega_{m}^2}{6\beta_{\rm DGP}^3(a)a^6}\left(1 - \frac{(\vec{k_1}\cdot \vec{k_2})^2}{k_1^2k_2^2}\right)\comma
\end{align}
For this model, and likely for Galileons in general, the $\gamma_2$ terms have the same $k_1,k_2$ dependence as in $\Lambda$CDM so the second order growth-factor becomes a function of time only. This means that it behaves just as $\Lambda$CDM albeit with different growth-factors.

\section{Including screening in modified gravity theories}\label{sec:screening}

One of the main ingredients of a successful modified gravity theory is a screening mechanism \cite{2010arXiv1011.5909K} that hides the modifications of gravity in high-density regions. {\it N}-body simulations of models with screening (see e.g. \cite{2008PhRvD..78l3524O,2009PhRvD..79h3518S,2009PhRvD..80d3001S,2011PhRvD..83d4007Z}) have shown that it is crucial to include the screening effect to get accurate results, for example linear perturbation theory might predict a $50\%$ enhancement of the matter power-spectrum relative to $\Lambda$CDM at some scale while simulations on the other hand might only show deviations at the few $\%$ level.

In \cite{2015PhRvD..91l3507W} a simplified approximative method to include screening was proposed which relies on combining spherically symmetric  analytical or semi-analytical solutions for the screening effect with a linear field equation. In effect it estimates from the amplitude of the density-field, the gravitational potential or it's gradient (depending on the model in question) how much of the mass contributes to the fifth-force and then uses this to correct the linearized field-equation. The linearized field equation can be rapidly solved using Fourier transforms instead of using a time consuming relaxation method to solve a highly non-linear field equation with bad convergence properties, as is done in most modified gravity {\it N}-body codes today.

\subsection{$f(R)$ gravity}
For $f(R)$ gravity \cite{2010LRR....13....3D}, which has the chameleon screening mechanism \cite{2004PhRvD..69d4026K}, we have that the fifth-force on an object (ignoring for now the finite range of the force) is given approximately by
\begin{align}
\vec{F}_\phi = \frac{1}{3}\cdot \vec{F}_{\rm Newton} \cdot \epsilon_{\rm screen}(\Phi_N)\comma
\end{align}
where
\begin{align}
 \epsilon_{\rm screen}(\Phi_N) = \text{Min}\left[1,\left|\frac{3f_{R}(a)}{2\Phi_N}\right|\right]\comma
\end{align}
and $\Phi_N$ is the standard Newtonian gravitational potential. The linearized field-equation on the other hand is given by
\begin{align}
\nabla_{\bf x}^2\phi = a^2m^2(a)\phi + \frac{1}{3}\cdot \BetaFac\delta\comma
\end{align}
where $m(a) = \frac{1}{3f_{RR}}$ is a model dependent function describing the inverse range of the fifth-force on cosmological scales \changes{and $\phi$ is related to $f_R$ via $\phi \equiv -\frac{1}{2}\log(f_R+1) \simeq -\frac{f_R}{2}$}. The field is normalized here such that $\vec{\nabla_{\bf x}}\phi$ corresponds to the fifth-force (i.e. the total force is $\vec{\nabla_{\bf x}}\Phi_N + \vec{\nabla_{\bf x}}\phi$). To include the effects of screening we solve the linear field equation
\begin{align}
\nabla_{\bf x}^2\phi = a^2m^2(a)\phi + \frac{1}{3}\cdot \BetaFac\delta \cdot  \epsilon_{\rm screen}(\Phi_N)\comma
\end{align}
in our simulation. $\Phi_N$ is easily computed from the density field which allows us to quickly solve for the effects of the fifth-force using Fourier transforms. This method allows us to perform modified gravity simulations at a computational cost that is not much larger ($20-50\%$ is a reasonable estimate) than for $\Lambda$CDM.

\subsection{nDGP}
For the normal-branch DGP model \cite{2000PhLB..485..208D,2010LRR....13....5M} with a $\Lambda$CDM background expansion, the modifications to the Poisson equation are given by $\Phi = \Phi_N + \phi$ where the scalar field $\phi$ is determined by
\begin{align}\label{eq:ndgpfieldeq}
\nabla_{\bf x}^2\phi + \frac{2r_c^2}{a^4}\left((\nabla_{\bf x}^2\phi)^2 - (\nabla_{\textbf{x}_i}\nabla_{\textbf{x}_j}\phi)^2\right) = \frac{\BetaFac\delta}{3\beta_{\rm DGP}(a)}\period
\end{align}
This equation is solved in modified gravity {\it N}-body simulations of this model.

For spherically symmetrical mass distributions the solution for the force $\vec{F}_\phi = \vec{\nabla_{\bf x}}\phi$ is given by
\begin{align}
\vec{F}_\phi = \frac{1}{3\beta_{\rm DGP}(a)}\cdot \vec{F}_{\rm Newton}\cdot  \epsilon_{\rm screen}(\rho)\comma
\end{align}
where
\begin{align}
 \epsilon_{\rm screen}(\rho) &= \frac{2\sqrt{1 + x}}{x}\comma\\
x &= \frac{8(r_cH_0)^2\Omega_m}{9\beta_{\rm DGP}^2(a)}\frac{\rho}{\overline{\rho}}\comma
\end{align}
where $\rho$ is the average density within a given radius. From this we can make the approximate linear field equation
\begin{align}
\nabla_{\bf x}^2\phi = \frac{1}{3\beta_{\rm DGP}(a)}\cdot \BetaFac\delta\cdot  \epsilon_{\rm screen}(\rho)\comma
\end{align}
which can be solved in the code to give the fifth-force. One problem with this equation is that the screening factor depends on density which means that the result will depend on the resolution of the simulation. To get around this issue we first smooth the density field with a Gaussian filter of a given radius $R$ ($R \sim 1 \text{Mpc}/h$ works well in practice) and use the smoothed density field to compute the screening factor above. \changes{This choice is motivated by the fact that the screening (Vainshtein) radius for the nDGP models we consider here is $\mathcal{O}(1) \text{Mpc}/h$ for typical halos we expect to have in our simulations. We have verified that the exact value of the smoothing radius does not significantly change our results by comparing the results we find for $R = 0.5, 1$ and $2$ Mpc$/h$.}

\section{Results}\label{sec:results}

In this section we show test runs of our code for some example models.

To start with we made sure the code is working correctly by performing some simple tests. First we use the scale-dependent solver to solve for $\Lambda$CDM and compare to the standard {\tt{L-PICOLA}} code. The agreement is found to be excellent ($\ll \%$ accuracy on all scales for $P(k)$).

\begin{figure*}
\begin{center}
\includegraphics[width=0.9\textwidth]{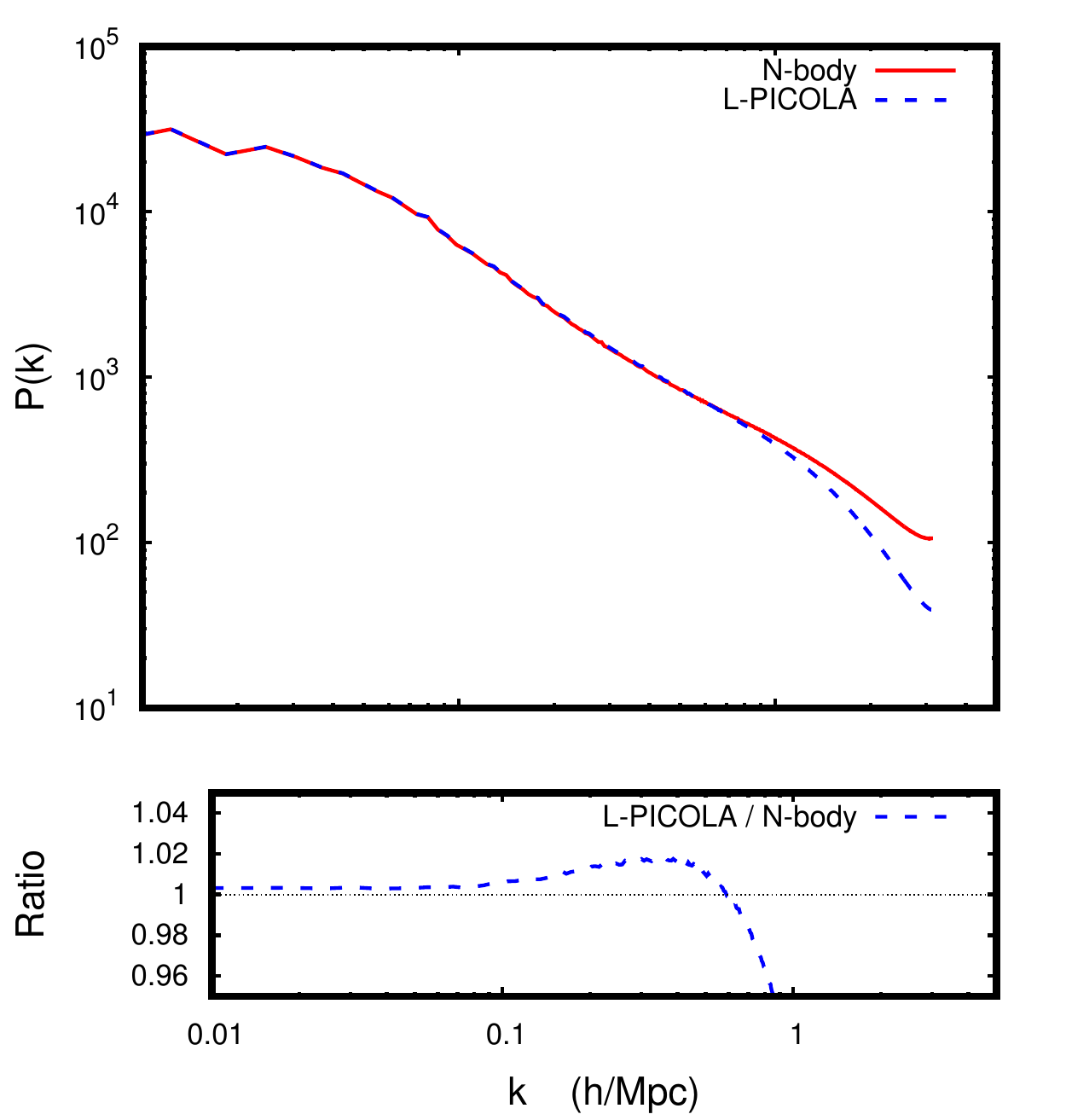}
\caption{The matter power-spectrum at redshift $z=0$ obtained from {\tt{L-PICOLA}} using a fixed mesh with $N = 1024^3$ gridcells in a box of size $B = 1024$ Mpc$/h$ and using $n = 30$ time steps compared to a high-resolution {\it N}-body simulation ({\tt{RAMSES}}) using the same initial conditions.}
\label{fig:ratiopofk}
\end{center}
\end{figure*}

Below we show comparisons of our code with results from true {\it N}-body simulations. To do this we created a module that reads in initial conditions from a given simulation and uses this to generate the displacement-fields which allow us to do a comparison without cosmic variance. In Fig.~\ref{fig:ratiopofk} we show a comparison of $P(k)$ for $\Lambda$CDM using {\tt{L-PICOLA}} (with $n = 30$ time steps) compared to the results of the {\it N}-body code {\tt{RAMSES}} \cite{2002A&A...385..337T}. The agreement is excellent on large scales, while for wavenumbers larger than $\sim k_{\rm Nyquist}/4 \sim 0.7 h/$Mpc the results starts to deviate as we cannot resolve smaller scales. In the rest of this paper we show the results relative to $\Lambda$CDM for runs with modified gravity models.

The (friend-of-friend) halo finder used in the analysis below is {\tt{MatchMaker}}\footnote{https://github.com/damonge/MatchMaker} and it was run with the linking-length $b = 0.2$. \finalchanges{The errors bars in the mass function plots are Poisson errors. Since the simulations were started form the same initial conditions these errors should be considered an upper limit to the shot noise and that it is likely significantly smaller than that.}

\subsection{$f(R)$ gravity}

The {\it N}-body simulation suite we used to test the $f(R)$ result of our code is taken from the modified gravity code comparison project \cite{2015MNRAS.454.4208W} (run with the {\tt{ISIS}} code \cite{2014A&A...562A..78L}) and consists of a $N = 512^3$ particle simulation in a $B = 250 $Mpc$/h$ box with a cosmology defined by $\Omega_m = 0.269$, $h=0.704$, $n_s=0.966$ and $\sigma_8=0.8$. The two $f(R)$ models have $|f_{R0}| = 10^{-5}$ (F5) and $|f_{R0}| = 10^{-6}$ (F6). The $f(R)$ simulations were run with the same initial condition as the $\Lambda$CDM simulation.

\begin{figure*}
\begin{center}
\includegraphics[width=0.9\textwidth]{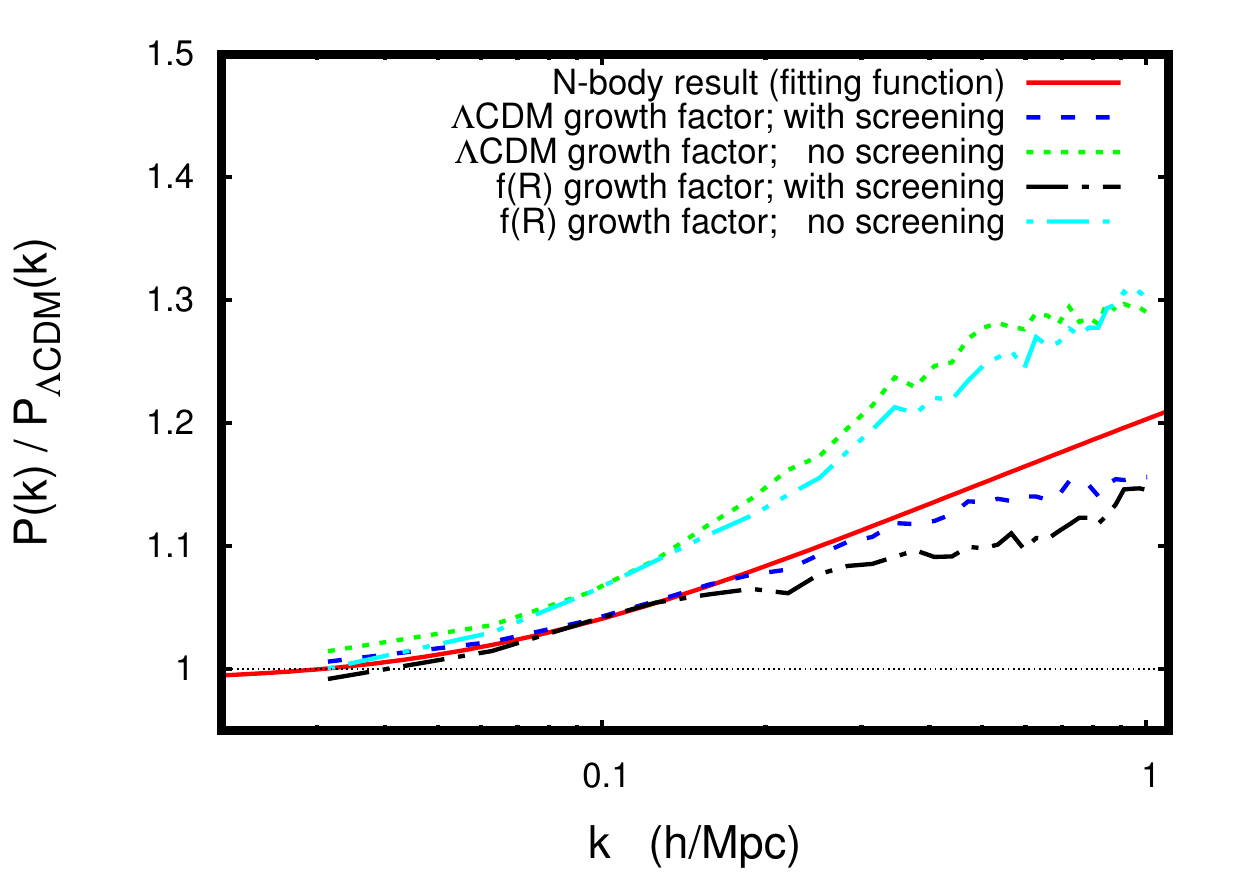}
\caption{The ratio of the matter power-spectrum in $f(R)$ to that in $\Lambda$CDM at redshift $z=0$ when using the true growth-factor(s) or using the $\Lambda$CDM ones plus the effect of including the screening method. Here we have used $n=10$ time steps.}
\label{fig:fofrgrowthfaccomp}
\end{center}
\end{figure*}

In Fig.~\ref{fig:fofrgrowthfaccomp} we show a comparison of the result we get when using the true $f(R)$ growth-factor versus using the $\Lambda$CDM growth-factor in the simulations. For this plot we have used $n=10$ time steps in the COLA simulations and we see a small difference in the power-spectrum at $z=0$. For $n>20$ the results are pretty much indistinguishable which happens because the more time steps we take the less effect the COLA approximation has on the final results. For a small number of time steps the COLA approximation is more important and the difference in the results comes from the true growth-factor taking some screening into account leading to a small reduction in power on non-linear scales. We also see that we significantly overestimate the true power-spectrum if we don't take screening of the fifth-force into account.

\begin{figure*}
\begin{center}
\includegraphics[width=0.9\textwidth]{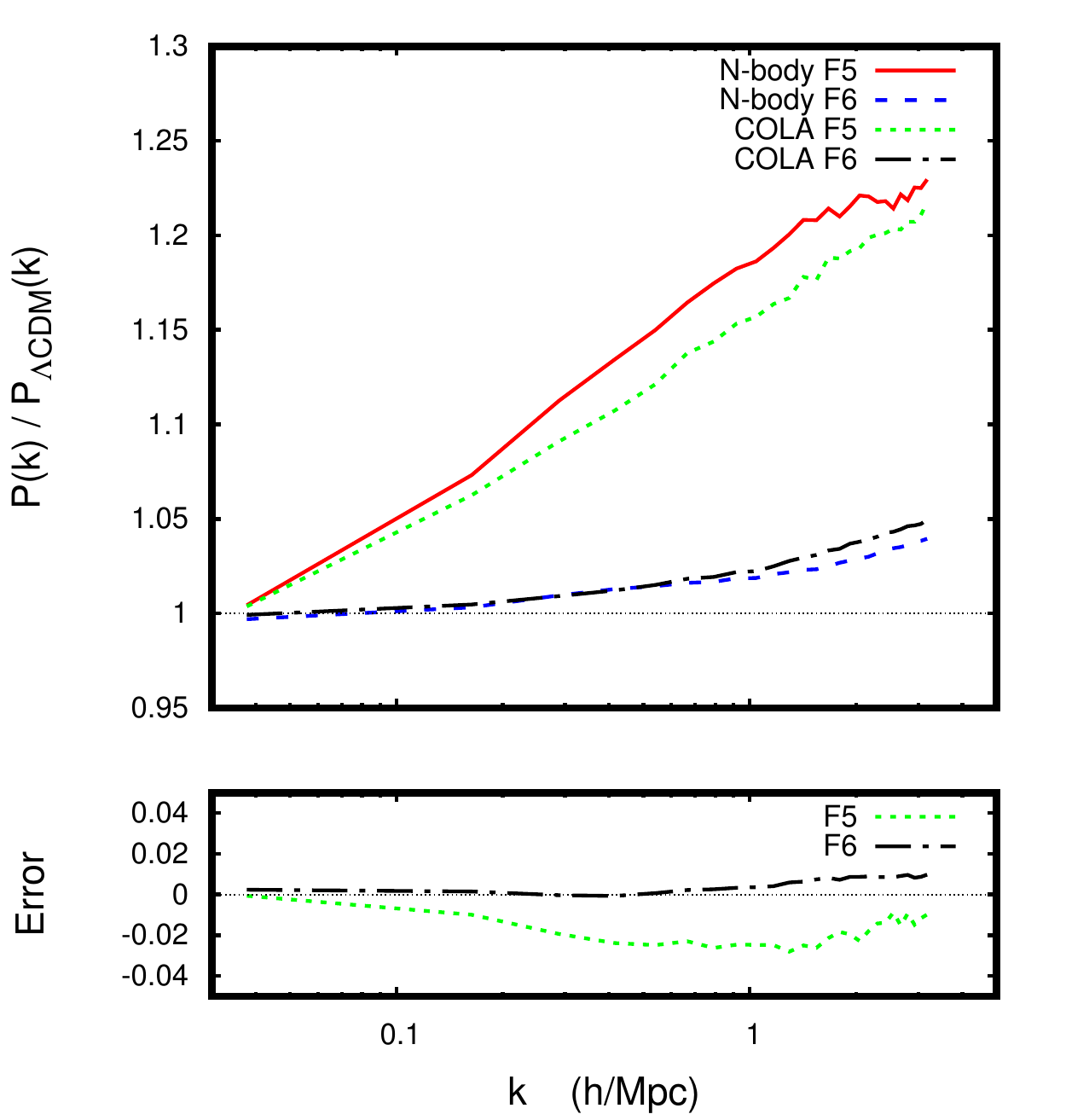}
\caption{The ratio of the matter power-spectrum in $f(R)$ to that in $\Lambda$CDM at redshift $z=0$. All simulations have been performed using the same initial conditions \changes{and we have used $n=30$ time-steps in the COLA simulations}. The {\it N}-body results correspond to modified gravity simulations solving the exact equations to get the fifth-force. For the COLA simulations we used the $\Lambda$CDM growth-factor. \changes{The lower panel shows $(P_{f(R)}/P_{\Lambda\rm CDM})^{\rm COLA} / (P_{f(R)}/P_{\Lambda\rm CDM})^{\rm {\it N}-body}$ - 1.}}
\label{fig:fofrpofk}
\end{center}
\end{figure*}

In Fig.~\ref{fig:fofrpofk} we show the fractional difference in the matter power-spectrum for $f(R)$ with respect to $\Lambda$CDM for our simulations including screening compared to the results of full {\it N}-body simulations. The agreement is $\lesssim 2$\% for F5 and $< 1\%$ for F6 up to $k\sim 3 h/$Mpc.

\begin{figure*}
\begin{center}
\includegraphics[width=0.9\textwidth]{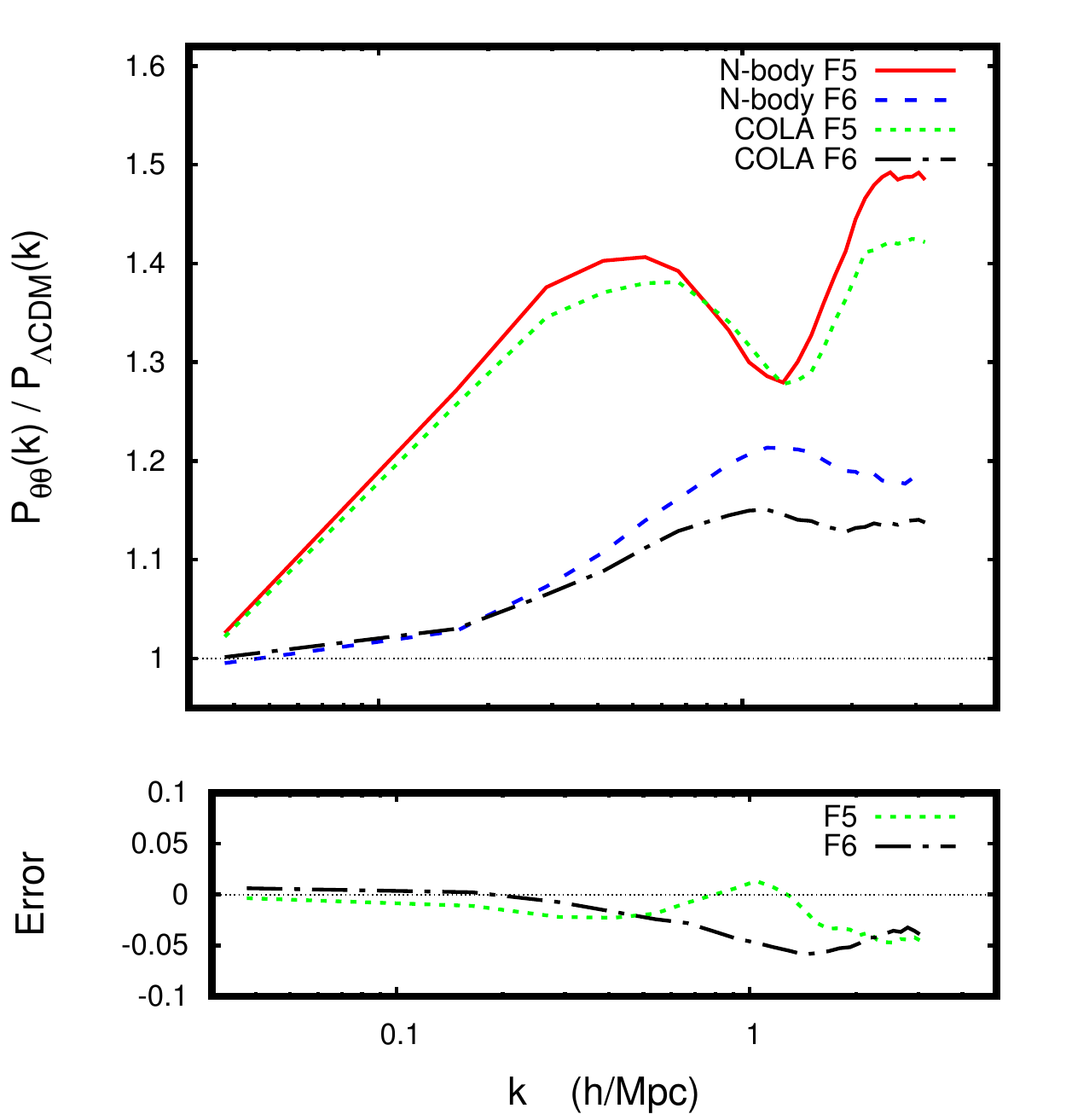}
\caption{The ratio of the velocity divergence power-spectrum in $f(R)$ to that in $\Lambda$CDM at redshift $z=0$. All simulations have been performed using the same initial conditions. The {\it N}-body results correspond to modified gravity simulations solving the exact equations to get the fifth-force. For the COLA simulations we used the $\Lambda$CDM growth-factor \changes{and $n=30$ time-steps. The lower panel shows $(P_{f(R)}/P_{\Lambda\rm CDM})^{\rm COLA} / (P_{f(R)}/P_{\Lambda\rm CDM})^{\rm {\it N}-body}$ - 1.}}
\label{fig:fofrtheta}
\end{center}
\end{figure*}

In Fig.~\ref{fig:fofrtheta} we show the fractional difference in the velocity divergence power-spectrum. The agreement is slightly worse than for the matter power-spectrum with up to $5\%$ deviation for F5 and up to $8\%$ for F6. This is still a decent agreement compared to the enhancement with respect to $\Lambda$CDM which is up to $\sim 50\%$ for F5 and up to $\sim 30\%$ for F6.

\begin{figure*}
\begin{center}
\includegraphics[width=0.9\textwidth]{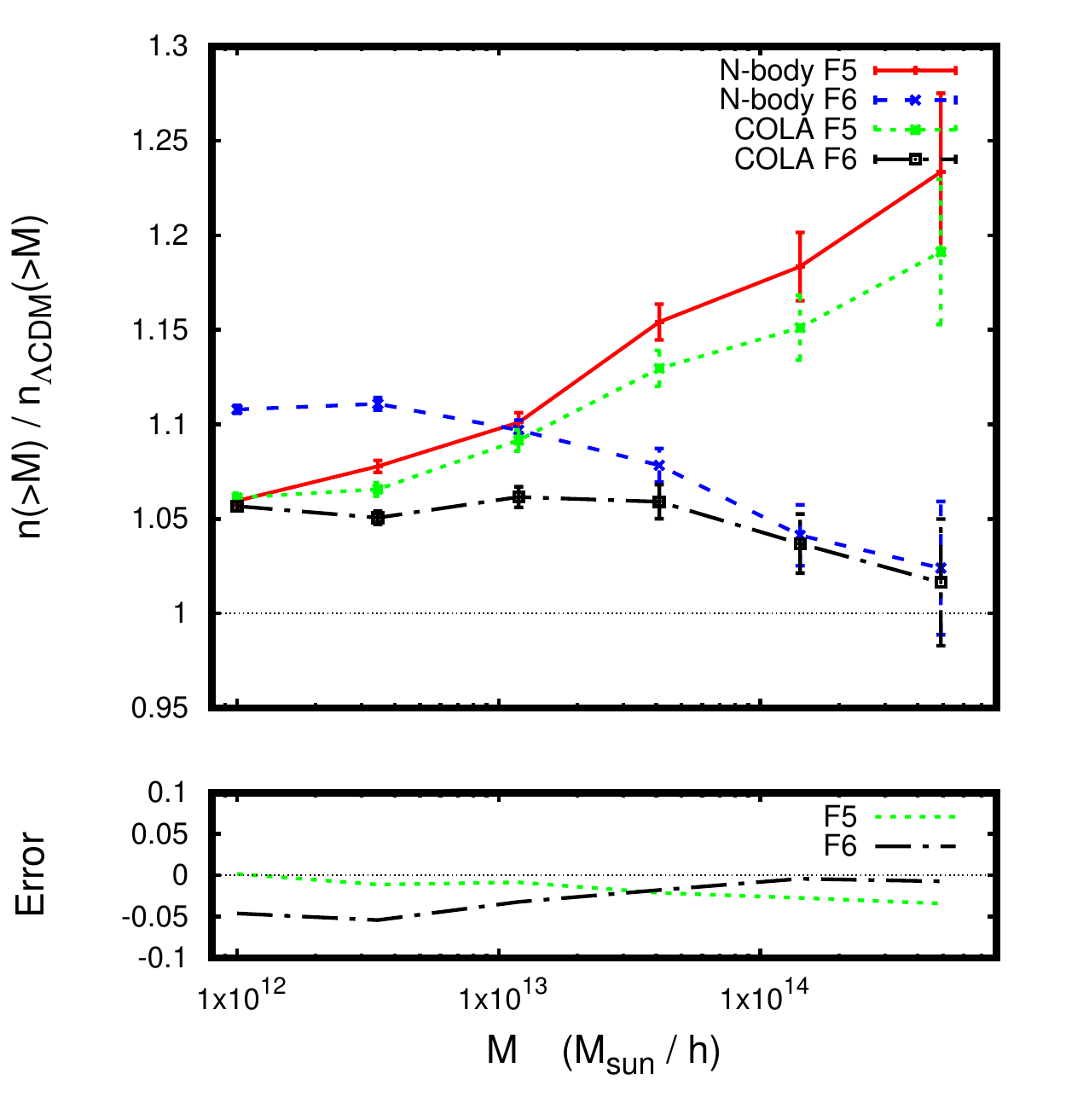}
\caption{The ratio of the halo mass function at $z=0$ in $f(R)$ to that in $\Lambda$CDM. All simulations have been performed using the same initial conditions. The {\it N}-body results correspond to modified gravity simulations solving the exact equations to get the fifth-force. For the COLA simulations we used the $\Lambda$CDM growth-factor \changes{and $n=30$ time-steps}. The error bars for the halo mass function are Poisson errors. \changes{The lower panel shows $(n_{f(R)}/n_{\Lambda\rm CDM})^{\rm COLA} / (n_{f(R)}/n_{\Lambda\rm CDM})^{\rm {\it N}-body}$ - 1.}}
\label{fig:fofrnofm}
\end{center}
\end{figure*}

In Fig.~\ref{fig:fofrnofm} we show the fractional difference in the halo mass function with respect to $\Lambda$CDM. The agreement is $\lesssim 2\%$ for all of the mass-range for F5, but for F6 we underestimate the enhancement of the mass function by approximately $5\%$ for $M \lesssim 5\cdot 10^{13}M_{\odot}/h$. This is the same as was found when using the screening method in full {\it N}-body simulations \cite{2015PhRvD..91l3507W} and this can therefore be attributed to this approximation. 

\subsection{nDGP}

The {\it N}-body simulation suite we used to test the nDGP version of our code was taken from \cite{2017arXiv170202348B} and was run with the {\tt{ECOSMOG}} code \cite{2012JCAP...01..051L}. The simulations have $N = 1024^3$ particles in a $B = 1024$ Mpc$/h$ box with a WMAP9 cosmology defined by $\Omega_m = 0.281$, $h = 0.697$, and $n_s = 0.971$. The two nDGP simulations have $r_cH_0 = 0.75$ (nDGP2) and $r_cH_0 = 4.5$ (nDGP3). These values correspond to having the same value of $\sigma_8(z=0)$ as the $f(R)$ models F5 and F6. The nDPG simulations were run with the same initial conditions as the $\Lambda$CDM simulation.

\begin{figure*}
\begin{center}
\includegraphics[width=0.9\textwidth]{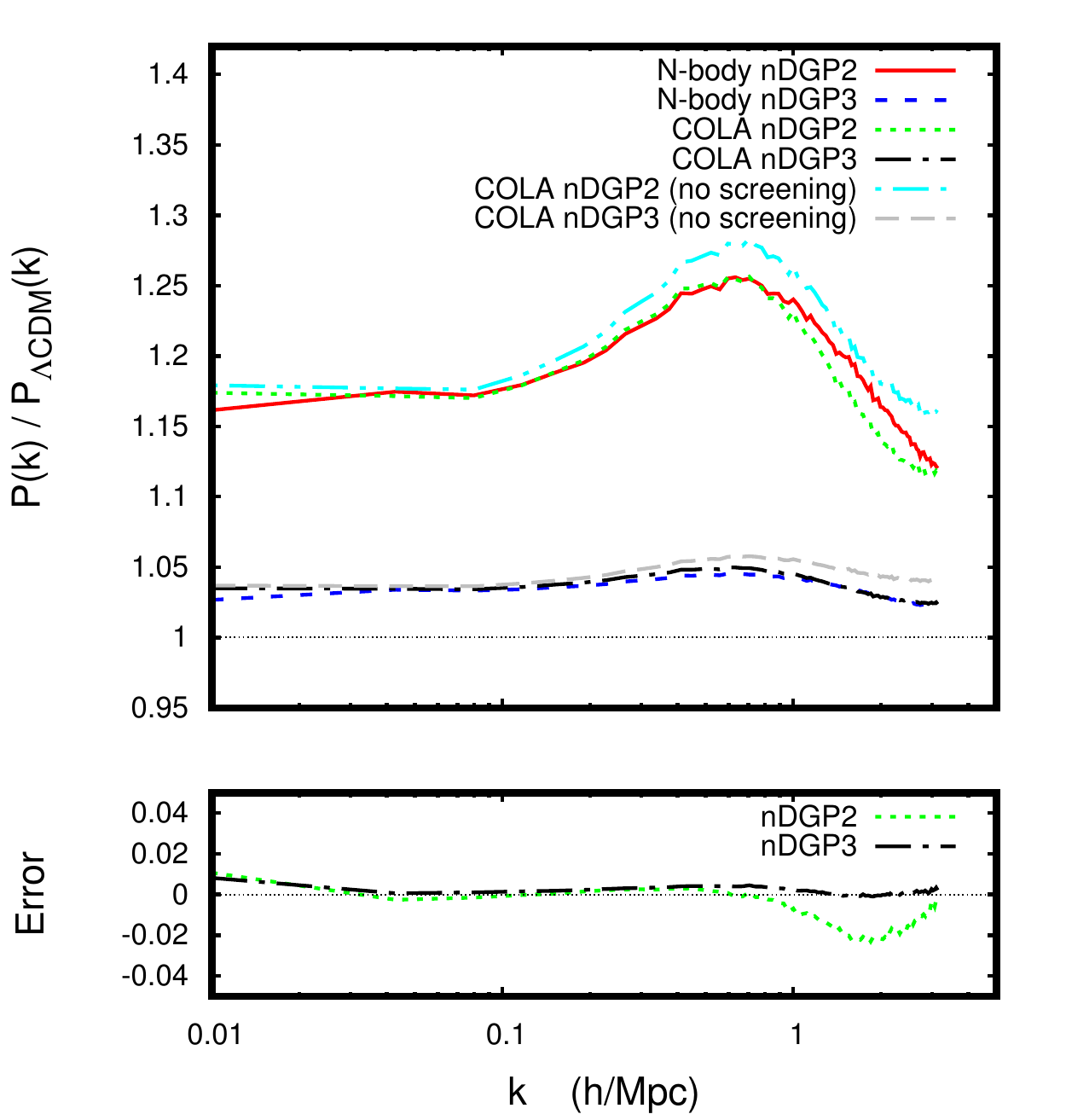}
\caption{The ratio of the matter power-spectrum at $z=0$ in nDGP to that in $\Lambda$CDM. All simulations have been performed using the same initial conditions. The {\it N}-body results correspond to modified gravity simulations solving the exact equations to get the fifth-force. For the COLA simulations we used \changes{$n=30$ time-steps and} a smoothing radius of $R = 1$ Mpc$/h$ to compute the screening factor for nDGP. \changes{The lower panel shows $(P_{\rm nDGP}/P_{\Lambda\rm CDM})^{\rm COLA} / (P_{\rm nDGP}/P_{\Lambda\rm CDM})^{\rm {\it N}-body}$ - 1.}}
\label{fig:dgppofk}
\end{center}
\end{figure*}

In Fig.~\ref{fig:dgppofk} we show the fractional difference in the matter power-spectrum for nDGP with respect to $\Lambda$CDM for our simulations with and without including screening compared to the results of full {\it N}-body simulations. The actual $P(k)$ starts to deviate from the {\it N}-body result already around $k\sim 0.5 h\text{Mpc}^{-1}$ while the enhancement has good $< 2\%$ accuracy all the way up to $k \sim 3h/$Mpc.

\begin{figure*}
\begin{center}
\includegraphics[width=0.9\textwidth]{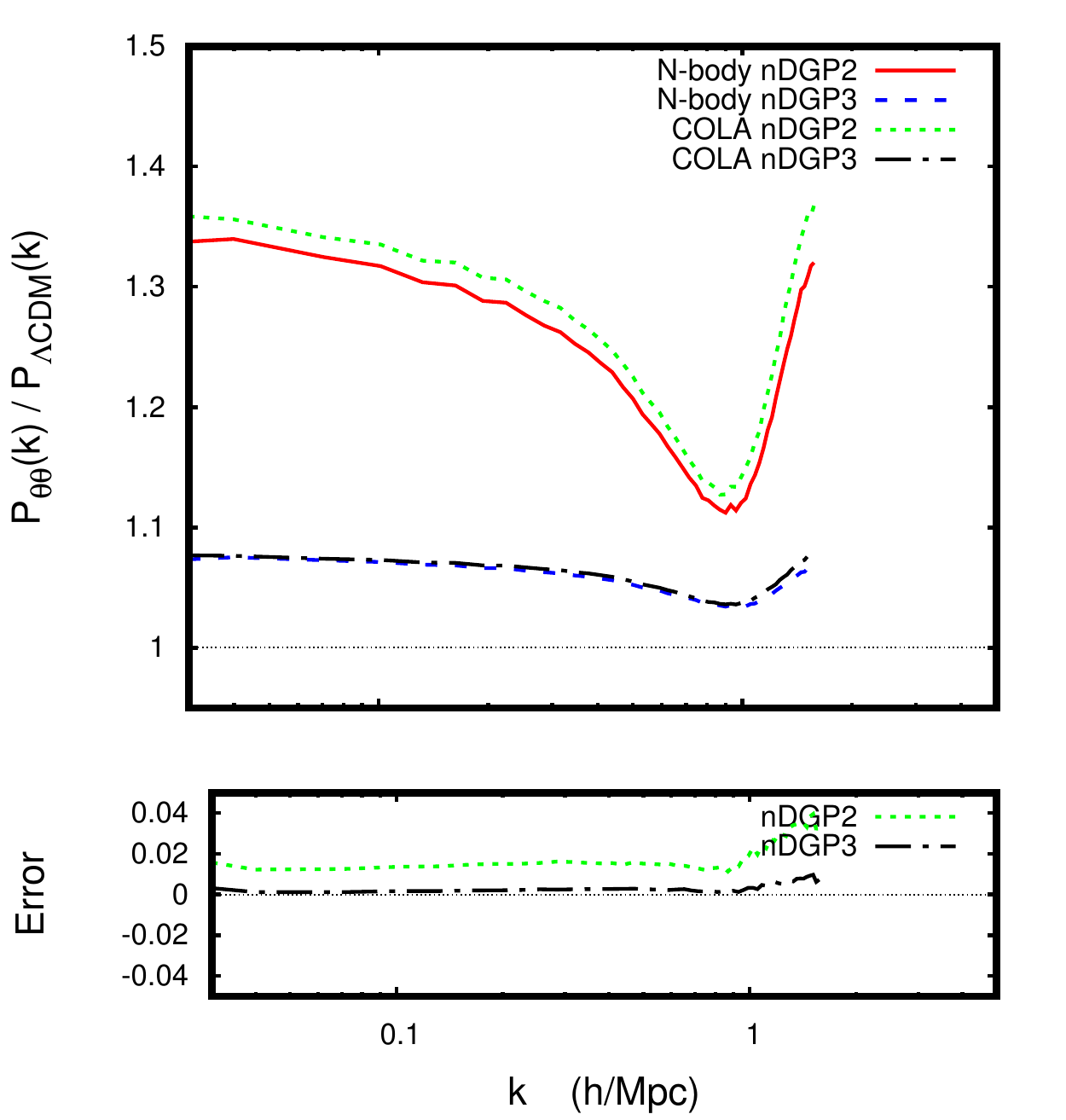}
\caption{The ratio of the velocity divergence power-spectrum in nDGP to that in $\Lambda$CDM at redshift $z=0$. All simulations have been performed using the same initial conditions \changes{and we have used $n=30$ time-steps in the COLA simulations}. The {\it N}-body results correspond to modified gravity simulations solving the exact equations to get the fifth-force. \changes{The lower panel shows $(P_{\rm nDGP}/P_{\Lambda\rm CDM})^{\rm COLA} / (P_{\rm nDGP}/P_{\Lambda\rm CDM})^{\rm {\it N}-body}$ - 1.}}
\label{fig:dgptheta}
\end{center}
\end{figure*}

In Fig.~\ref{fig:dgptheta} we show the fractional difference in the velocity divergence power-spectrum with respect to $\Lambda$CDM compared to the results of full {\it N}-body simulations. The agreement is $\lesssim 2$\% up to $k\sim 2h/$Mpc which is fairly small compared to the large signal relative to $\Lambda$CDM which is $\sim 7$\% and $\sim 20$\% for the two models respectively.

\begin{figure*}
\begin{center}
\includegraphics[width=0.9\textwidth]{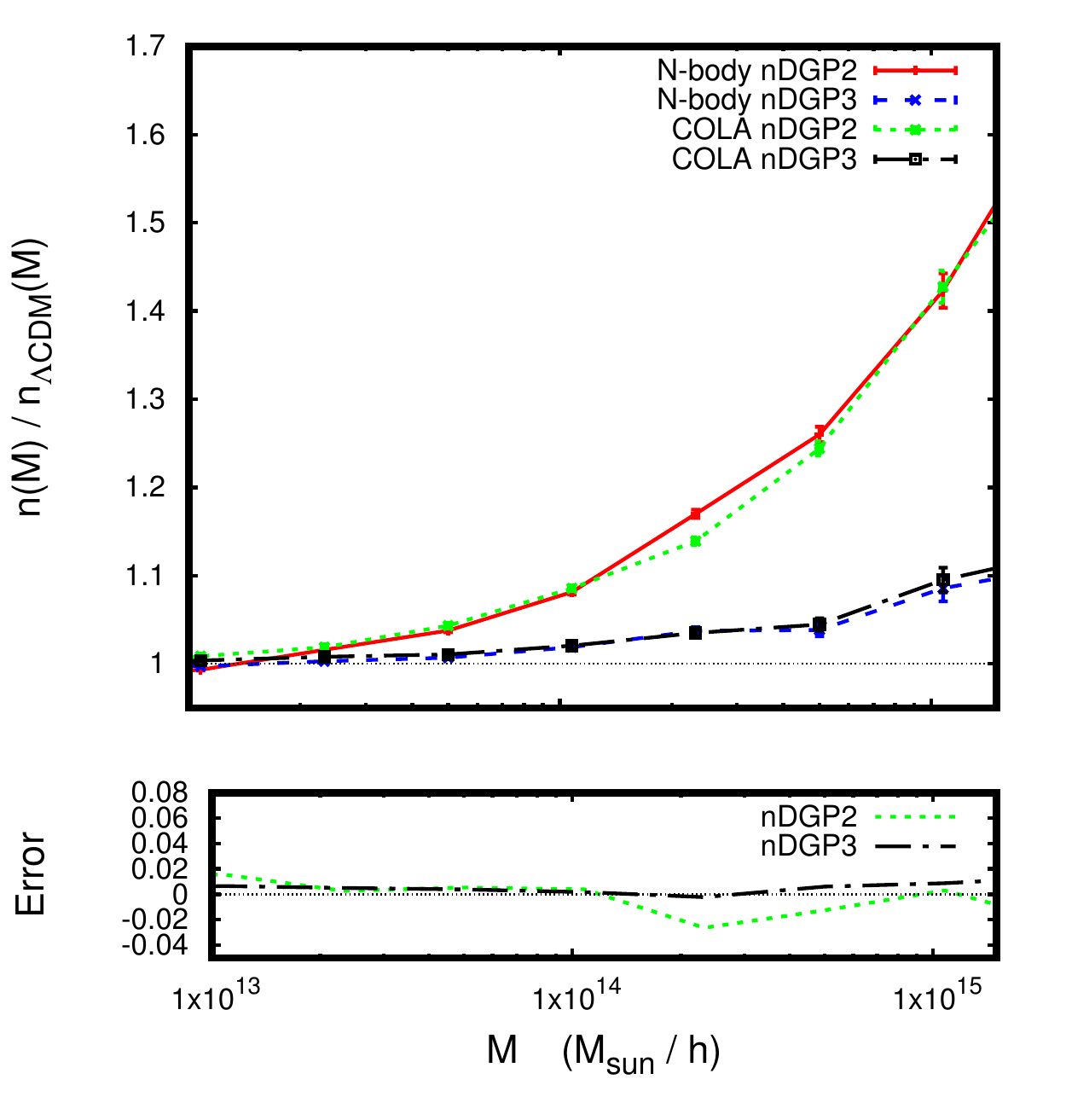}
\caption{The ratio of the halo mass function in nDGP to that in $\Lambda$CDM at redshift $z=0$. All simulations have been performed using the same initial conditions \changes{and we have used $n=30$ time-steps in the COLA simulations}. The {\it N}-body results correspond to modified gravity simulations solving the exact equations to get the fifth-force. For the COLA simulations we used the smoothing radius of $R = 1$ Mpc$/h$ to compute the screening factor for nDGP. The error bars for the halo mass function are Poisson errors. \changes{The lower panel shows $(n_{\rm nDGP}/n_{\Lambda\rm CDM})^{\rm COLA} / (n_{\rm nDGP}/n_{\Lambda\rm CDM})^{\rm {\it N}-body}$ - 1.}}
\label{fig:dgpnofm}
\end{center}
\end{figure*}

In Fig.~\ref{fig:dgpnofm} we show the fractional difference in the halo mass function with respect to $\Lambda$CDM. The agreement is $\lesssim 2\%$ for the entire mass-range $10^{12}-10^{15}M_{\odot}/h$ probed by this simulation.

The COLA approach for these types of models works nearly as well as for $\Lambda$CDM and the computational cost is only $\sim 30\%$ larger and comes from computing the smoothed density-field at every time-step which requires one additional Fourier transform.

\subsection{Dependence on the number of steps}

The run-time of the code is roughly proportional to the number of time steps so the fewer steps we can use the better.

\begin{figure*}
\begin{center}
\includegraphics[width=0.5\textwidth]{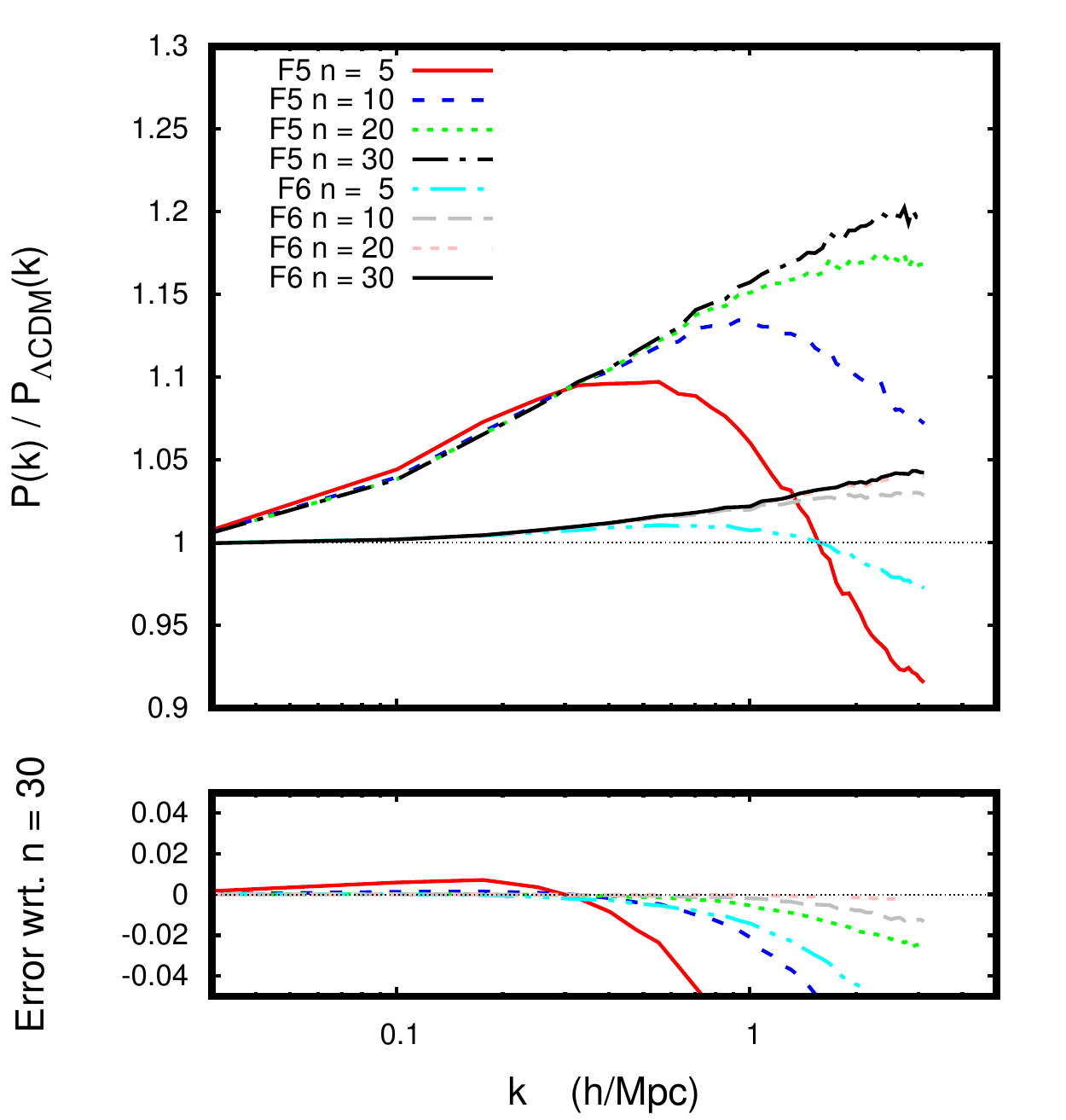}\includegraphics[width=0.5\textwidth]{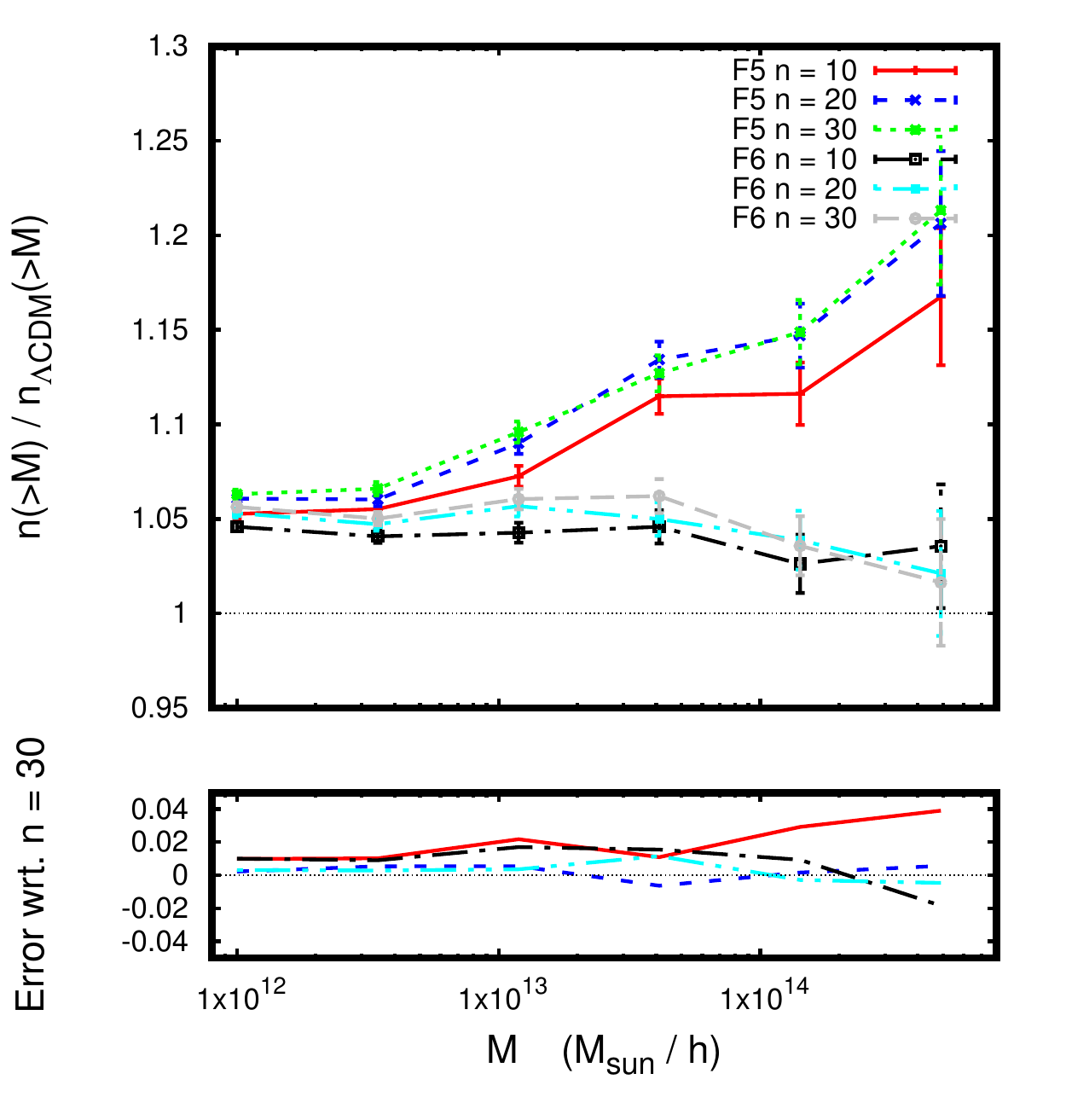}
\caption{The ratio of the matter power-spectrum (left) and halo mass function (right) in $f(R)$ to that in $\Lambda$CDM at redshift $z=0$ for the two $f(R)$ models F5 and F6 for different number of time steps. The ratio in each case is with respect to a $\Lambda$CDM simulation using the same number of steps. In the lower panel we show the fractional difference in the ratio with respect to the $n=30$ run. The error bars for the halo mass function are Poisson errors.}
\label{fig:fofrnstep}
\end{center}
\end{figure*}

In Fig.~\ref{fig:fofrnstep} we show how the results for the matter power-spectrum and halo mass function in our $f(R)$ simulations depend on the number of time steps. The enhancement of the power-spectrum relative to $\Lambda$CDM is seen to have converged for $k<1 h/$Mpc already when using $n = 10$ time steps for both models. To get a similar convergence on the smaller scales probed by our simulations we need to go up $\sim 20-30$ time steps. For the halo mass function we are within $5\%$ of the $n=30$ result across the whole mass range already at $n=10$ and for $n=20$ the results have practically converged.

\begin{figure*}
\begin{center}
\includegraphics[width=0.5\textwidth]{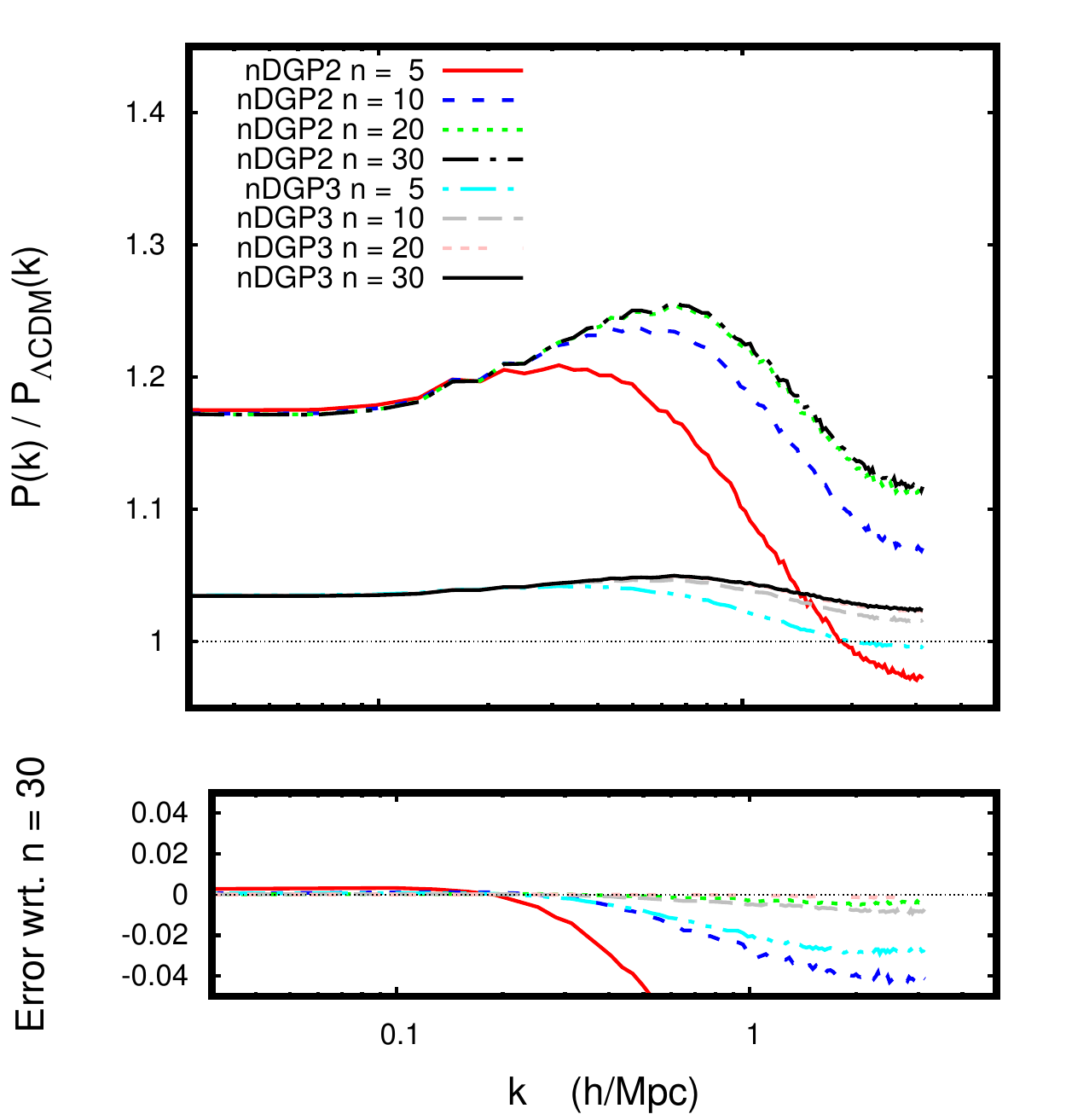}\includegraphics[width=0.5\textwidth]{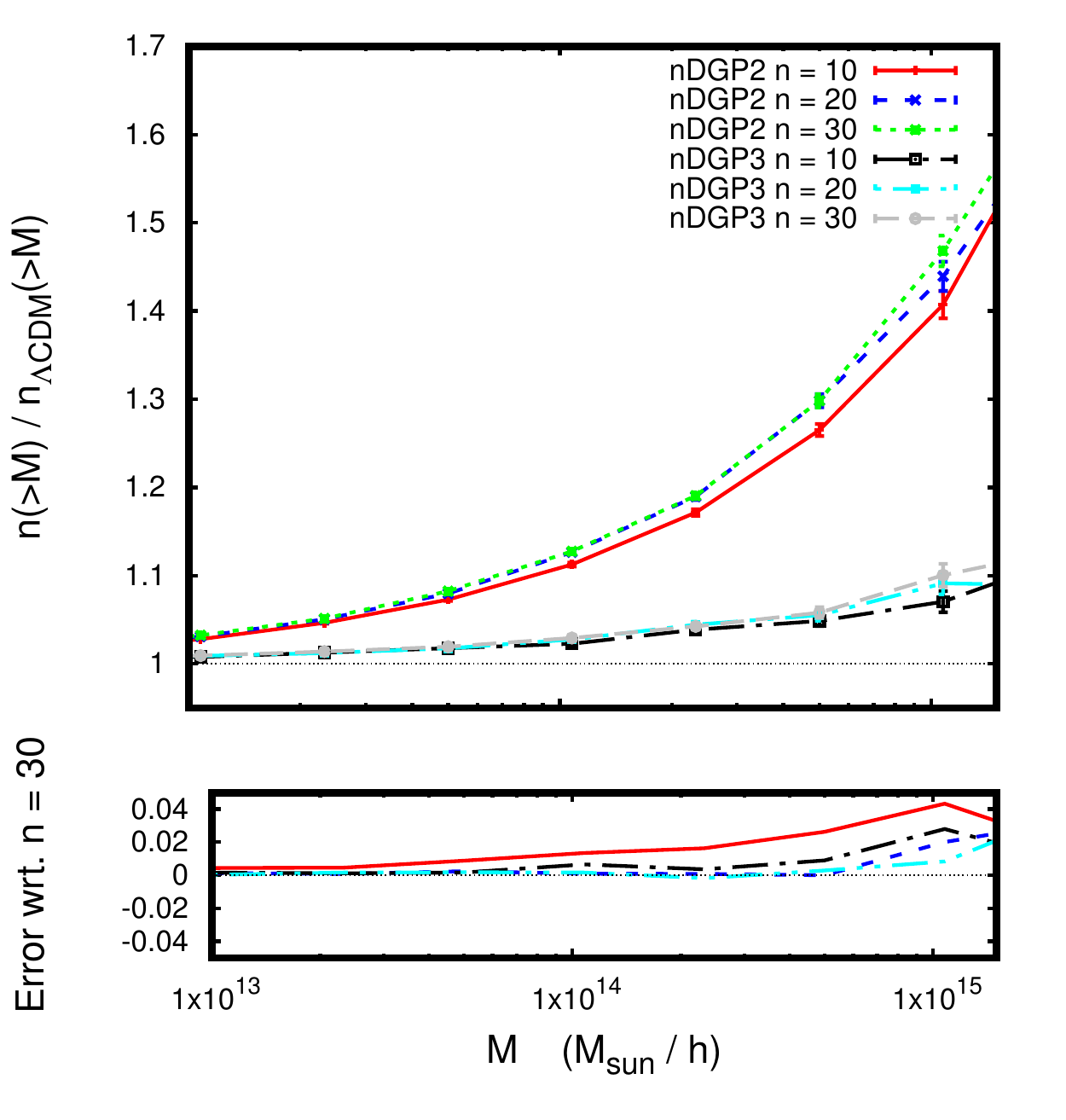}
\caption{The ratio of the matter power-spectrum (left) and halo mass function (right) in nDGP to that in $\Lambda$CDM at redshift $z=0$ for the two nDGP models nDGP2 and nDGP3 for different number of time steps. The ratio in each case is with respect to a $\Lambda$CDM simulation using the same number of steps. In the lower panel we show the fractional difference in the ratio with respect to the $n=30$ run. The error bars for the halo mass function are Poisson errors.}
\label{fig:dgpnstep}
\end{center}
\end{figure*}

In Fig.~\ref{fig:dgpnstep} we show the corresponding result for our nDGP simulations. The same type of behavior as we saw for $f(R)$ is also found here: $n = 10$ time steps is enough to get the power-spectrum boost-factor (ratio with respect to $\Lambda$CDM) correct to $\sim 2\%$ up to $k=1 h$/Mpc while to get full convergence we need $\sim 20$ time steps. The boost-factor for the halo mass function is within $4\%$ of the $n=30$ result in the $n=10$ run across the whole mass range.

These results show that we can get away with using a fairly low number of time steps $n\sim 10-20$ and still maintain percent level accuracy in the boost-factors.

\section{Conclusions}\label{sec:conclusion}

We have presented a code that uses the COLA approach to large scale structure formation. This code is applicable to a general class of cosmological models that exhibit scale dependent growth. The main focus here has been on modified gravity theories, but it is also possible to use this scheme to include other effects like massive neutrinos similar to what was done in \cite{2017JCAP...01..008R} for PINOCCHIO. \changes{This has recently been shown to work very well in \cite{2017arXiv170508165W}.} The code comes with a general implementation of an approximate method for including the three most common types of screening one finds in modified gravity theories. We have also implemented a general parameterization of scalar-tensor theories of the chameleon form using the $\{m(a),\beta(a)\}$ formulation together with commonly studied models like $f(R)$, DGP and Jordan-Brans-Dicke. \changes{Built in to the code is also tools for doing on-the-fly computation of (friend-of-friend) halo catalogs plus both real space and redshift space matter power-spectra.}

By comparing to full modified gravity simulations we have demonstrated that the approach works very well. The boost-factors $X / X_{\rm LCDM}$ for clustering statistics like power-spectra and halo mass function (both computed using COLA) are able to recover the true {\it N}-body result to percent level accuracy deep into the non-linear regime ($k\sim 3 h/$Mpc) even when using a low number of COLA time steps.

The addition of scale-dependent growth does have the drawback of slowing down the COLA approach relative to $\Lambda$CDM by a factor of \changes{$\sim 3-4$} in the current implementation, but as we have shown, and was previously found in \cite{2016arXiv161206469V}, for $f(R)$ (and likely other models of this form) one can get away with using the $\Lambda$CDM growth-factor making this approach only about $\sim 30\%$ slower than $\Lambda$CDM. However the scale-dependent implementation is still needed to verify this approximation and there is no guarantee it will hold for a general model. 

For the nDGP models we tested (which should also hold for Galileon models in general) the growth-factors remain scale-independent to second order and the only computational overlay for these simulations is in the computation of the screening factor which requires one extra Fourier transform per step making it only $\sim 30\%$ slower than the corresponding $\Lambda$CDM simulation. It should be straight forward to implement a general Galileon model using the results for the screening-function presented in \cite{2013JCAP...11..056B}.

\section{Acknowledgement}
We would like to thank Benjamin Bose for useful discussion and Wojciech Hellwing for providing us with a code to compute velocity divergence power-spectra \cite{2013MNRAS.428..743L}. \changes{We would also like to thank Alejandro Avil\'es and Jorge L. Cervantes-Cota for pointing out a missing term in our 2LPT scale-dependent growth-equation}. BSW is supported by the U.K. Science and Technology Facilities Council (STFC) research studentship. KK and HAW are supported by the European Research Council through 646702 (CosTesGrav). KK is also supported by the UK Science and Technologies Facilities Council grants ST/N000668/1. GBZ is supported by NSFC Grant No. 11673025, and by a Royal Society-Newton Advanced Fellowship.

\appendix

\section{Implementation details}\label{sec:implementation}

The main change we need to implement is to account for the scale-dependent growth-factors. This is easily done by storing the Fourier transform of the initial displacement-fields, multiplying by the growth-factors and performing a Fourier transform to get the real-space displacement-fields at every time-step. Having computed the displacement-fields we assign the displacement-vector $\vec{\Psi}(\vec{q},\tau)$ to the particles. This needs to be done at every step.

An additional complication comes when we run with several processors. The particles require the displacement-field at their original Lagrangian positions so for particles that have crossed a CPU boundary we need inter-CPU communication to obtain this. This is done by storing the original CPU-id and $q$-coordinate with each particle which requires $8\cdot N_{\rm particles}$ bytes of memory. Additional (temporary) memory is needed to store both $\frac{d\vec{\Psi}}{d\tau}$ and $\frac{d^2\vec{\Psi}}{d\tau^2}$ which adds another $12\cdot 4 = 48$ bytes per particle compared to $\Lambda$CDM.

Finally we also need extra memory to store the initial displacement-fields (in $k$-space), temporary memory to perform the Fourier-transforms, and temporary memory to compute the screening factor. This makes the scale-dependent implementation much more memory expensive that the standard $\Lambda$CDM implementation.

\section{Summary of the general equations solved by the code}
The fiducial choice for the background expansion is $\Lambda$CDM, however it is easy to modify this by redefining the function $H(a)$ and $\frac{dH(a)}{da}$.

For the linear perturbations the user must provide $\mu(k,a)$ (and possibly $\gamma_2$ if one has this available, otherwise put this to $0$). The growth factors are then determined by
\begin{align}
\frac{d^2D_1}{d\tau^2} - \BetaFac\mu(k,a) D_1 &= 0\comma\\
\frac{d^2D_2}{d\tau^2} - \BetaFac\mu(k,a) D_2 &= - \BetaFac \mu(k,a) D_1^2(k,a)\times\nonumber\\&\left(1 + \frac{2\gamma_2 a^4H^2}{\BetaFac \mu(k,a)}\right)\period
\end{align}
For the {\it N}-body part of the code we have implemented routines to solve any field equation of the form
\begin{align}
\nabla_{\bf x}^2\phi = m^2(a)a^2\phi + C(a)\cdot \BetaFac \delta \cdot \epsilon_{\rm screen}(\Phi_N,|\vec{\nabla_{\bf x}}\Phi_N|,\nabla_{\bf x}^2\Phi_N)\comma
\end{align}
where $\phi$ is normalized such that the total force on the particles is $\vec{\nabla_{\bf x}}\Phi_N + \vec{\nabla_{\bf x}}\phi$. This covers the three most widely known screening mechanisms: chameleon, k-Mouflage and Vainhstein. The user can pick any of these three screening methods (i.e. either screening by potential, gradient or density) and the screening-function $\epsilon_{\rm screen}$ needs to be specified.

For potential screening this is done automatically by the code (see next section) as long as the user specifies the two functions $m(a)$ and $\beta(a)$ (and in this case $C(a) = 2\beta^2(a)$).

For gradient (k-Mouflage) screening arising from $P(X = \frac{1}{2}(\nabla_{\bf x}\phi)^2)$ Lagrangians with a conformal coupling to matter of the form $e^{\frac{\beta\phi}{M_{\rm Pl}}}$ then the screening function is determined by
\begin{align}
P_X^2(X_*) X_* &= (2\beta M_{\rm Pl})^2 |\vec{\nabla_{\bf x}}\Phi_N|^2\comma\\
\epsilon_{\rm screen}(|\vec{\nabla_{\bf x}}\Phi_N|) &= \text{Min}\left[1,\frac{1}{P_X(X_*)}\right]\period
\end{align}
For these models we have $C(a) = 2\beta^2$ and the linear growth factor is determined by $\mu = \frac{2\beta^2}{P_X(X(a))}$ where $X(a)$ is the cosmological value of $X$. As no {\rm N}-body simulation of these types of models is found in the literature we have not yet tested this approach, but all of the methods needed have been included in the code and one only needs to provide an expression for $X_*(|\vec{\nabla_{\bf x}}\Phi_N|^2)$ and $X(a)$ to use it.

For Vainshtein screening (DGP, Galileon models) one needs to specify $\epsilon_{\rm screen}(\nabla_{\bf x}^2\Phi_N \propto \rho)$ and the coupling $C(a)$ which for nDGP is simply $C(a) = \frac{1}{3\beta_{\rm DGP}(a)}$ as shown in Eq.~(\ref{eq:ndgpfieldeq}). For these models we have $m(a) = 0$, i.e. the range of the fifth-force is infinite. Since the density is highly resolution dependent we need to use a smoothed density field to compute the screening. We have implemented three common choices for the Fourier space smoothing filter, namely the Gaussian, top-hat and sharp-$k$ window functions. The user only needs to choose a smoothing filter and a smoothing scale $R_{\rm smooth}$.

\section{Implementation of general $\{m(a),\beta(a)\}$ models}

As shown in \cite{2012PhRvD..86d4015B} a general scalar-tensor theory with a potential and a conformal coupling to matter that shows the screening effect is uniquely defined by specifying two time-dependent functions on the cosmological background: the coupling strength of the fifth-force $\beta(a)$ and the mass of the scale (inverse range of the fifth-force) $m(a)$. Given these functions we can reconstruct the potential $V(\phi)$ and the conformal coupling $A(\phi)$. Examples of models of this form are the chameleon, the symmetron, and the environmental dependent dilaton model. {\it N}-body simulations for several different functional forms of $m(a)$ and $\beta(a)$ were performed in \cite{2013JCAP...04..029B,2012JCAP...10..002B}.

Here we will describe the implementation of a general $\{m(a),\beta(a)\}$ model in our code. At the level of linear perturbations we have
\begin{align}
\mu(k,a) = 1 + 2\beta^2(a) \frac{k^2}{k^2 + a^2m^2(a)}\period
\end{align}
\finalchanges{
and to second order we have
\begin{align}
\gamma_2^{\rm E} = \frac{m^2(a) \frac{dm^2(a)}{da}\beta^2(a)\Omega_m}{2H_0^4\Pi(k)\Pi(k_1)\Pi(k_2)} \frac{k^2}{a^4H^2}\comma
\end{align}
where $\Pi(k) = \left(\frac{k}{aH_0}\right)^2 + \frac{m^2(a)}{H_0^2}$. } For the {\it N}-body part the field equation reads
\begin{align}
\nabla_{\bf x}^2\phi = m^2(a)a^2\phi + 2\beta^2(a) \cdot 4\pi G \delta\rho \cdot \epsilon_{\rm screen}(\Phi_N)\comma
\end{align}
where $\phi$ is normalized such that $\vec{\nabla_{\bf x}}\Phi_N + \vec{\nabla_{\bf x}}\phi$ is the total force on the particles. The screening function is given by
\begin{align}
\epsilon_{\rm screen}(\Phi_N) = \text{Min}\left[1,\left|\frac{\Phi_{\rm crit}(a)}{\Phi_N}\right|\right]\comma
\end{align}
where the critical potential for screening is
\begin{align}
\Phi_{\rm crit}(a) = \Phi_{\rm crit}(a_{\rm ini})  + \frac{9\Omega_m}{2\beta(a)}\int_{a_{\rm ini}}^a \frac{\beta(a')}{\frac{m^2(a')}{H_0^2}a^4}{\rm d}a'\period
\end{align}
The code solves the integral above for $\Phi_{\rm crit}(a)$, however if analytical expressions are available then it's recommended to use these instead.

For example the ($n=1$) Hu-Sawicky $f(R)$ model can be recast of this form with
\begin{align}
\beta(a) &= \frac{1}{\sqrt{6}} \comma\\
m^2(a) &= H_0^2\frac{\Omega_m + 4\Omega_\Lambda}{2|f_{R0}|}\left(\frac{\Omega_m a^{-3} + 4\Omega_\Lambda}{\Omega_m + 4\Omega_\Lambda}\right)^{3}\comma
\end{align}
and the integral above gives rise to (in the limit $a_{\rm ini} \to 0$)
\begin{align}
\Phi_{\rm crit}(a) = \frac{3f_{R0}}{2}\left(\frac{\Omega_m a^{-3} + 4\Omega_\Lambda}{\Omega_m + 4\Omega_\Lambda}\right)^{2}\period
\end{align}
Another example is the symmetron model for which
\begin{align}
\beta(a) &= \beta_*\sqrt{1 - \frac{a_*^3}{a^3}}\comma\\
m^2(a) &= m_*^2\left(1 - \frac{a_*^3}{a^3}\right)\comma
\end{align}
where $\beta_*,\frac{m_*}{H_0},a_*$ are dimensionless parameters and we take $\beta(a) = m(a) = 0$ if $a < a_*$. The critical screening value becomes (we put $a_{\rm ini} = a_*$ as the fifth-force is not active for $a < a_*$)
\begin{align}
\Phi_{\rm crit}(a) = \frac{3\Omega_m}{2a_*^3}\frac{H_0^2}{m_*^2}\period
\end{align}
In this simple formulation we have ignored the additional screening effect in high density regions coming from the fact that $\beta(\phi) \to 0$ as the ambient density gets larger and larger. This illustrates how easy it is to include a new model of this form.

\bibliographystyle{JHEP}
\bibliography{references}
\end{document}